\documentclass[onecolumn,prl, preprint, superscriptaddress]{revtex4-2}
\usepackage{bm}
\usepackage[colorlinks=true,linkcolor=blue,citecolor=blue]{hyperref}
\usepackage{times}
\usepackage{amsmath}
\usepackage{amssymb}
\usepackage{amsthm}
\usepackage{amsfonts}
\usepackage{enumerate}
\usepackage{latexsym}
\usepackage{ifpdf}
\usepackage{natbib}
\usepackage{psfrag}
\newcommand{\beq}{\begin{equation}}
\newcommand{\eeq}{\end{equation}}
\usepackage{graphicx}
\usepackage{makeidx}
\usepackage{color}
\usepackage{array}
\usepackage{multirow}
\usepackage{siunitx}
\usepackage{}
\usepackage{tabularx}
\begin{document}


\title{Janus-faced influence of oxygen vacancy in  high entropy oxide films with Mott electrons}

\author {Suresh Chandra Joshi}
\affiliation {Department of Physics, Indian Institute of Science, Bengaluru 560012, India}
\author {Nandana Bhattacharya}
\affiliation {Department of Physics, Indian Institute of Science, Bengaluru 560012, India}
\author {Manav Beniwal}
\affiliation {Department of Physics, Indian Institute of Science, Bengaluru 560012, India}
\author {Jyotirmay Maity}
\affiliation {Department of Physics, Indian Institute of Science, Bengaluru 560012, India}
\author {Prithwijit Mandal}
\affiliation {Department of Physics, Indian Institute of Science, Bengaluru 560012, India}
\author{Hua Zhou}
\affiliation {Advanced Photon Source, Argonne National Laboratory, Lemont, Illinois 60439, USA}
\author {Christoph Klewe}
\affiliation {Advanced Light Source, Lawrence Berkeley National
Laboratory, Berkeley, California 94720, USA}
\author {Srimanta Middey}
 \email {smiddey@iisc.ac.in}
\affiliation {Department of Physics, Indian Institute of Science, Bengaluru 560012, India}


\begin{abstract}

Contrary to traditional approaches, high entropy oxides (HEOs) strategically employ cationic disorder to engineer tunable functionalities. This disorder, stemming from multiple elements at the same crystallographic site, disrupts local symmetry and induces local distortions. By examining a series of single-crystalline [La$_{0.2}$Pr$_{0.2}$Nd$_{0.2}$Sm$_{0.2}$Eu$_{0.2}$]NiO$_{3-\delta}$ thin films, we demonstrate herein that the creation of oxygen vacancies (OVs) further offers a powerful means of tailoring electronic behavior of HEOs by concurrently introducing disorder in the oxygen sublattice and doping electrons into the system.
Increasing OV concentration leads to a monotonic increase in room-temperature sheet resistance. A striking feature is the Janus-faced response of the metal-insulator transition (MIT) to OVs due to the interplay among correlation energy scales, electron doping, and disorder. Unlike the monotonous influence of OV observed for the MIT in VO$_2$ and V$_2$O$_3$, initial OV doping lowers the MIT temperature here, whereas higher OV levels completely suppress the metallic phase. Magnetotransport measurements further reveal weak localization, strong localization as a function of $\delta$. Moreover, the disorder on both $RE$ and oxygen sublattices is responsible for the Mott-Anderson insulator state.  These findings surpass the scope of the recently featured `electron antidoping' effect and demonstrate the promising opportunity to utilize OV engineering of HEOs for Mottronics and optoelectronics applications.   
       
\end{abstract}

\maketitle

High-entropy oxides (HEOs), comprising five or more elements occupying the same crystallographic site, have emerged as a promising class of materials with immense potential for both fundamental research, and electronic and energy applications ~\cite{Yoo:2024p466474,Tsai:2023p2302979,Zou:2024p3449234530,Oses:2020P295309,Sivak:2025p216101}. The inherent cation disorder in HEOs alters electronic interactions, offering significant advantages over conventional oxides~\cite{Mazza:2024p230501,Oses:2020P295309,Xiong:2024p2415351}.
This unique characteristic gives rise to a diverse range of physical phenomena, including nanoscale polarization~\cite{Xiong:2024p2415351}, exchange bias~\cite{Mazza:2022p2200391,Zou:2024p3449234530}, electron-phonon decoupling~\cite{Zheng:2024p7650}, and spin state transitions with concomitant modulation of carrier type~\cite{Zhang:2024p2406885}, etc. These intriguing properties translate into promising applications such as energy storage~\cite{Zhang:2024p185189,Yang:2022p1074-1080}, neuromorphic computing~\cite{Yoo:2024p466474,Tsai:2023p2302979}, and electrocatalysis~\cite{Sarkar:2019p1806236,Patel:2023p031407}, etc.  While disorder is widely recognized as a driving factor for various electronic phenomena such as weak localization, Anderson localization, many-body localization, electron glass, etc, in conventional oxides~\cite{Anderson:1958p1492,Amir:2011p235,Abanin:2019p021001}, these effects are rarely explored in HEOs, primarily due to their very low electrical conductivity.

Notably, while the cation sublattice in HEOs is intrinsically disordered, the oxygen sublattice has generally remained ordered, with only minor displacements of oxygen anions to accommodate local structural distortions. This work explores a largely uncharted strategy—introducing disorder within the oxygen sublattice—as an effective approach to unlock quantum transport phenomena in HEOs. As a model system to demonstrate this concept, we focus on [La$_{0.2}$Pr$_{0.2}$Nd$_{0.2}$Sm$_{0.2}$Eu$_{0.2}$]NiO$_{3}$ [(LPNSE)NO], an exceptionally rare HEO exhibiting a temperature-dependent metal-insulator transition (MIT)~\cite{Patel:2020p071601}. Rare-earth nickelates are characterized by a negative effective charge transfer energy ($\Delta$), where the ground state electronic wavefunction is dominated by the $d^8\underline{L}$ configuration, alongside a $d^7$ contribution (Ni$^{3+}$: 3$d^7$ and $\underline{L}$ denotes a hole on oxygen)~\cite{Middey:2016p305334}. The MIT in these systems is understood as a bond disproportionation transition ($d^8\underline{L}$ +$d^8\underline{L}$ $\rightarrow$ $d^8$+$d^8\underline{L}^2$) involving a significant rearrangement of the charge in the oxygen sublattice~\cite{Park:2012p156402,Bisogni:2016p13017,Middey:2018p156801}. The MIT has been also linked with polaron condensation~\cite{Shamblin:2018p86}.
Unlike band semiconductors like Si, where carrier addition enhances electrical conductivity, recent studies on $RE$NiO$_3$ have shown that electron doping actually decreases conductivity~\cite{Shi:2014p4860,Kotiuga:2019p21992,Li:2021p187602}. This effect, termed as `electron antidoping', has been associated with the annihilation of the polaron~\cite{Liu:2019p106403,Shi:2024p256502}.

Given the generally insulating/semiconducting nature of most HEOs~\cite{Aamlid:2023p5991}, (LPNSE)NO provides an ideal platform to investigate electronic transport through controlled OV creation, as  OVs also add electrons into the system, apart from introducing disorder. This study involved the growth of [La$_{0.2}$Pr$_{0.2}$Nd$_{0.2}$Sm$_{0.2}$Eu$_{0.2}$]NiO$_{3-\delta}$ films with a range of oxygen deficiencies ($\delta$) via pulsed laser deposition (PLD). The preservation of the perovskite structure in all films, as verified by synchrotron X-ray diffraction, indicates a random distribution of OVs. X-ray absorption spectroscopy (XAS) revealed that the $\delta$ varied between 0 and 0.2 across our film series and a weakening of the Ni $d$-O $p$ hybridization strength as $\delta$ increased. 
A key finding of our study is the Janus-faced influence of OVs achieved through our synthesis method, a stark departure from the monotonous OV effects observed in other first-order MIT materials like VO$_2$, and V$_2$O$_3$ and electron antidoping effect in $RE$NiO$_3$~\cite{Guo:2021p72,Jeong:2013p14021405,Park:2020P1401,Brockman:2011p152105,Shi:2014p4860,Kotiuga:2019p21992,Li:2021p187602}.
Initially, increasing OVs lowered the MIT temperature in present case, but further increases drove the system towards a Mott-Anderson-like insulating state via an intermediate weak localization phase. 
This study demonstrates that introducing disorder within the oxygen sublattice offers a powerful means to control the functional properties of HEOs.

  \begin{figure}[]
    \centering
    \includegraphics[width= .8\textwidth]{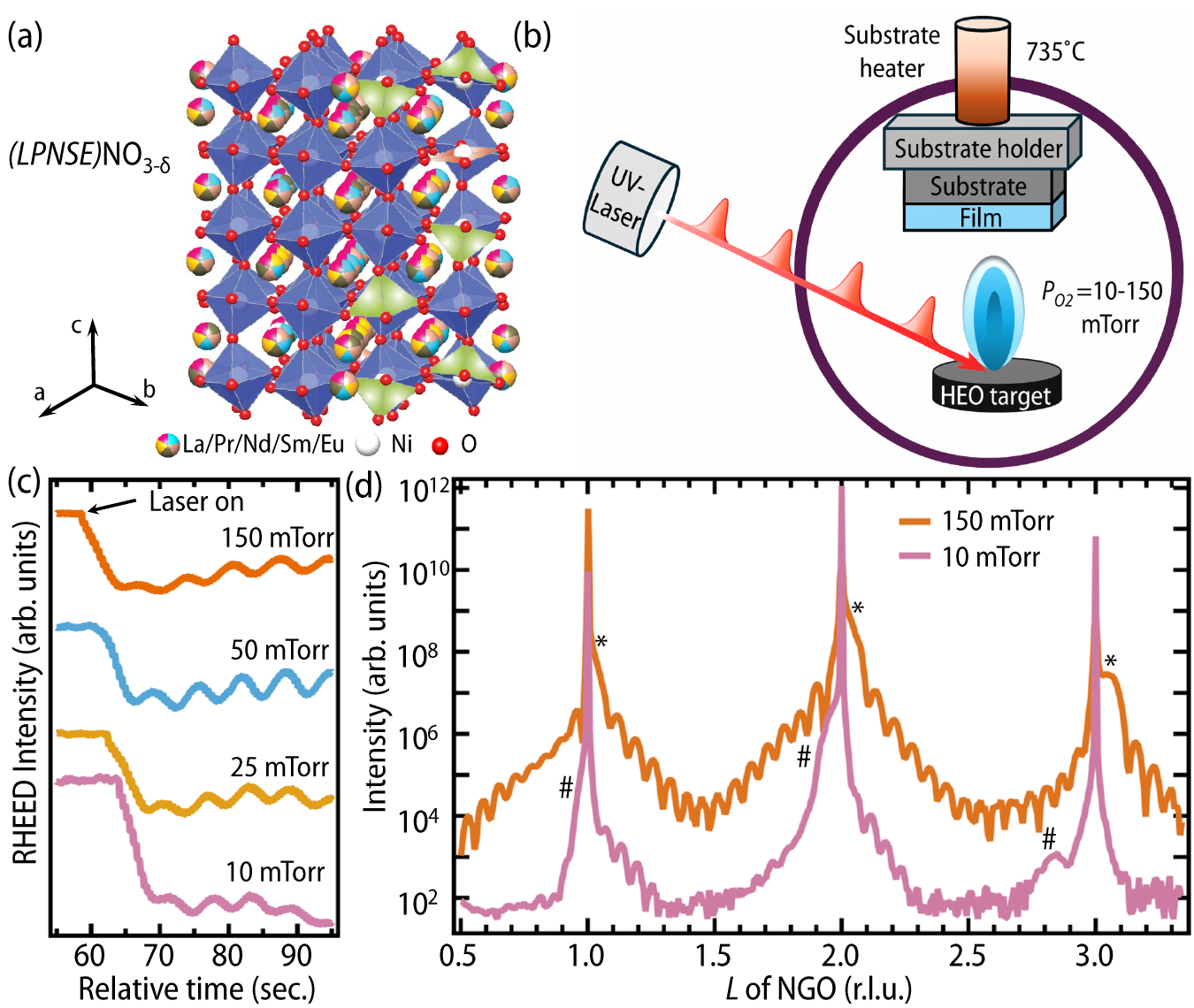}
    \caption{(a) Oxygen deficient (LPNSE)NO$_{3-\delta}$ structure. (b) Schematic diagram for PLD process, where  $P_{O_2}$ denotes the partial oxygen pressure, used during the deposition. (c) RHEED intensity oscillation during the growth for first five uc of several films. (d) Synchrotron XRD along (0 0 $L$) Bragg rod for the 150 and 10 mTorr samples. The reciprocal lattice unit (r.l.u.) has been defined with respect to the NGO. }
    \label{Fig1}
\end{figure}


While bulk $RE$NiO$_3$ necessitates high-pressure synthesis to stabilize Ni$^{3+}$, epitaxy with the single crystalline substrate can stabilize this oxidation state in thin film geometry without the need for elevated oxygen pressures~\cite{Gorbenko:2002p4026}. A series of single crystalline thin films of (LaPrNdSmEu)$_{0.2}$NiO$_{3-\delta}$ with thickness $\sim$ 5-6 nm were grown on NdGaO$_3$  [0 0 1]$_\mathrm{pc}$ (NGO) substrate by a pulsed laser deposition system. To vary the oxygen content systematically, films were grown with a systematic variation in dynamic oxygen pressure ($P_{O_{2}}$), ranging from 150 mTorr to 10 mTorr [see Fig.~\ref{Fig1}(b)]. A reflection high energy electron diffraction (RHEED) imaging system, connected with the deposition chamber, was used to monitor the progress of the growths. RHEED intensity oscillation was observed for the growth of the first 10-15 unit cells [Fig.~\ref{Fig1}(c)], and all films investigated in this manuscript are 15 uc thick. For each growth, the substrate temperature during the deposition was  735$^\circ$C, and the samples were cooled down to room temperature at the same dynamic $P_{O_{2}}$ without further annealing.
The presence of similar streaky patterns in specular and off-specular spots in RHEED images across all films provides compelling evidence for retention of the overall perovskite phase with smooth surface morphology for RHEED and atomic force microscopy image). Synchrotron X-ray diffraction scan along the (0 0 $L$)$_\mathrm{pc}$ Bragg rod revealed film peaks [marked by $\star$ for 150 and \# for 10 mTorr sample] and sharp substrate peaks [Fig.~\ref{Fig1} (d)], confirming single crystalline nature of the films growth along with the preservation of the overall perovskite phase. The observation of Laue oscillations adjacent to film peaks further testifies high crystalline quality. The peak position also shifts to a lower $L$ value with the lowering of $P_{O_2}$ as expected for lattice expansion due to OVs.

To verify the introduction of OVs and examine consequent changes in the electronic structure of these films with the decrease in $P_{O_{2}}$, we performed XAS measurements at the Eu $M_{4,5}$, Ni $L_{3,2}$ and O $K$-edges at 300 K. These measurements were conducted in total electron yield (TEY) mode at beamline 4.0.2 of the Advanced Light Source (ALS), USA.  Given the effective probing depth of TEY mode is approximately 9-10 nm, our XAS measurements provide information about the electronic structure across the entire film thickness. 
Unlike the other rare-earth ($RE$: La, Nd, Pr, Sm) ions, which exhibit exclusively +3 oxidation state, Eu can exist in both +2 and +3 states~\cite{Wei:2023peadh3327}. The measurement of  Eu $M_{4,5}$ edge XAS for the samples grown at 150 mTorr and 10 mTorr indeed confirm that Eu remains in the +3 oxidation state.

\begin{figure}[]
    \centering
    \includegraphics[width= .8\textwidth]{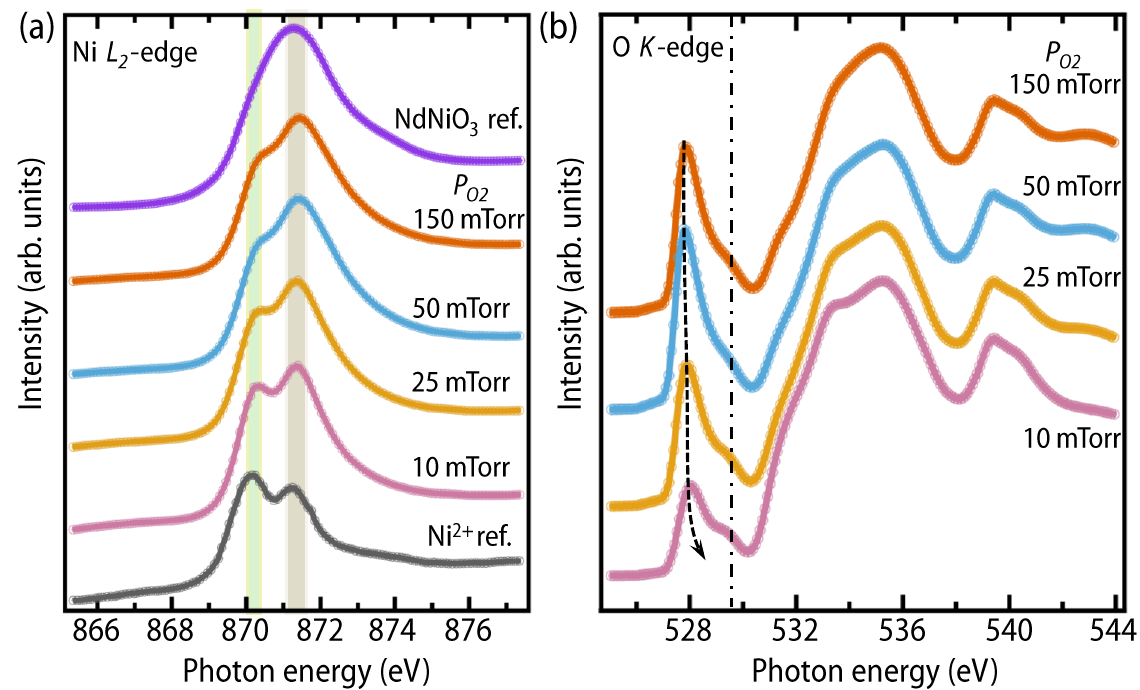}
    \caption{ XAS measured at 300 K for (a) the Ni $L_2$-edge, with NiO and NdNiO$_3$ included as references, and (b) the O $K$-edge for films, grown under different $P_{O_2}$.  In (b) the dashed arrow (around 528 eV) indicates the shift in pre peak position while the dotted line highlights the  appearance of prominent shoulder peak upon decreasing $P_\mathrm{O2}$.  All these spectra are normalized to zero at pre-edge and unity at the post-edge, and are vertically shifted for clarity.}
    \label{Fig:2}
\end{figure}

In Fig. \ref{Fig:2}(a), we show  the Ni $L_2$ edge XAS spectra [La $M_4$ edge overlaps strongly with the Ni $L_3$ edge.
The spectra for metallic NdNi$^{3+}$O$_3$ and insulating Ni$^{2+}$O have also been plotted for easy comparison. 
For the film deposited at 150 mTorr, a minor spectral feature around 870 eV is observed, which may be attributed to the presence of a small quantity of Ni$^{2+}$ and nanoscale chemical inhomogeneity at the $RE$ site~\cite{Bhattacharya:2025p2418490}. The spectral evolution in the other films is analyzed relative to this  film. A progressive increase in the Ni$^{2+}$ feature is observed in the spectra of films grown at lower $P_{O_{2}}$. By analyzing these spectra as a linear combination of NiO and 150 mTorr sample spectra, we estimate $\delta$ to be around 0.015, 0.14, 0.2 for the 50 mTorr, 25 mTorr and 10 mTorr samples, respectively. Our X-ray photoelectron spectroscopy measurements also confirmed similar $\delta$  for the 10 mTorr sample.
Overall, our doping levels are substantially lower than those in recent reports on heavily electron-doped $RE$NiO$_3$~\cite{Kotiuga:2019p21992,Li:2021p187602}, allowing us to access an unexplored regime.

The presence of OV should significantly influence the hybridization of Ni-O bonds, which has been further examined by O $K$-edge XAS [see Fig.~\ref{Fig:2}(b)]. The pre-peak observed at 527.8 eV for the 150 mTorr film [Fig.\ref{Fig:2}(b)]  arises due to to the 3$d^8\underline{L}\rightarrow\underline{c}3d^8$ transition for the Ni$^{3+}$ in octahedral coordination [$\underline{c}$ denotes a core hole in oxygen 1$s$ state]~\cite{Middey:2016p305334}. The broad feature around 535 eV corresponds to $RE$ 5$d$ states, whereas the spectral features around 540-542 eV is related to Ni 4$sp$ states, all hybridized with O 2$p$~\cite{Abbate:2002p155101}.  The spectrum for the 50 mTorr sample displayed comparable features, consistent with similar OV content found by Ni XAS. However, further reduction in $P_{O_{2}}$
  led to significant spectral changes within the 526-534 eV region. Notably, the pre-peak intensity at 527.8 eV diminishes, testifying a substantial reduction in Ni-O hybridization as anticipated in the presence of OVs~\cite{Middey:2014p6819}. Interestingly, this pre-peak also shifts to higher photon energy with the decrease $P_{O_{2}}$, which can be attributed to a decrease in charge transfer energy ($\Delta$)~\cite{Chakhalian:2011p116805,Liu:2013p2714,Middey:2018p045115}.
  Additionally, a spectral feature develops around 529.4 eV, attributed to Ni in a square planar arrangement~\cite{Abbate:2002p155101,Horiba:2007p155104}. The intensity of this feature increases with increasing $\delta$. Overall, our XRD and XAS measurements revealed OVs dope electrons into the system and leads to a random distribution of Ni with reduced coordination, while the overall perovskite lattice structure is preserved.

\begin{figure}[]
    \centering  \includegraphics[width=0.8\textwidth]{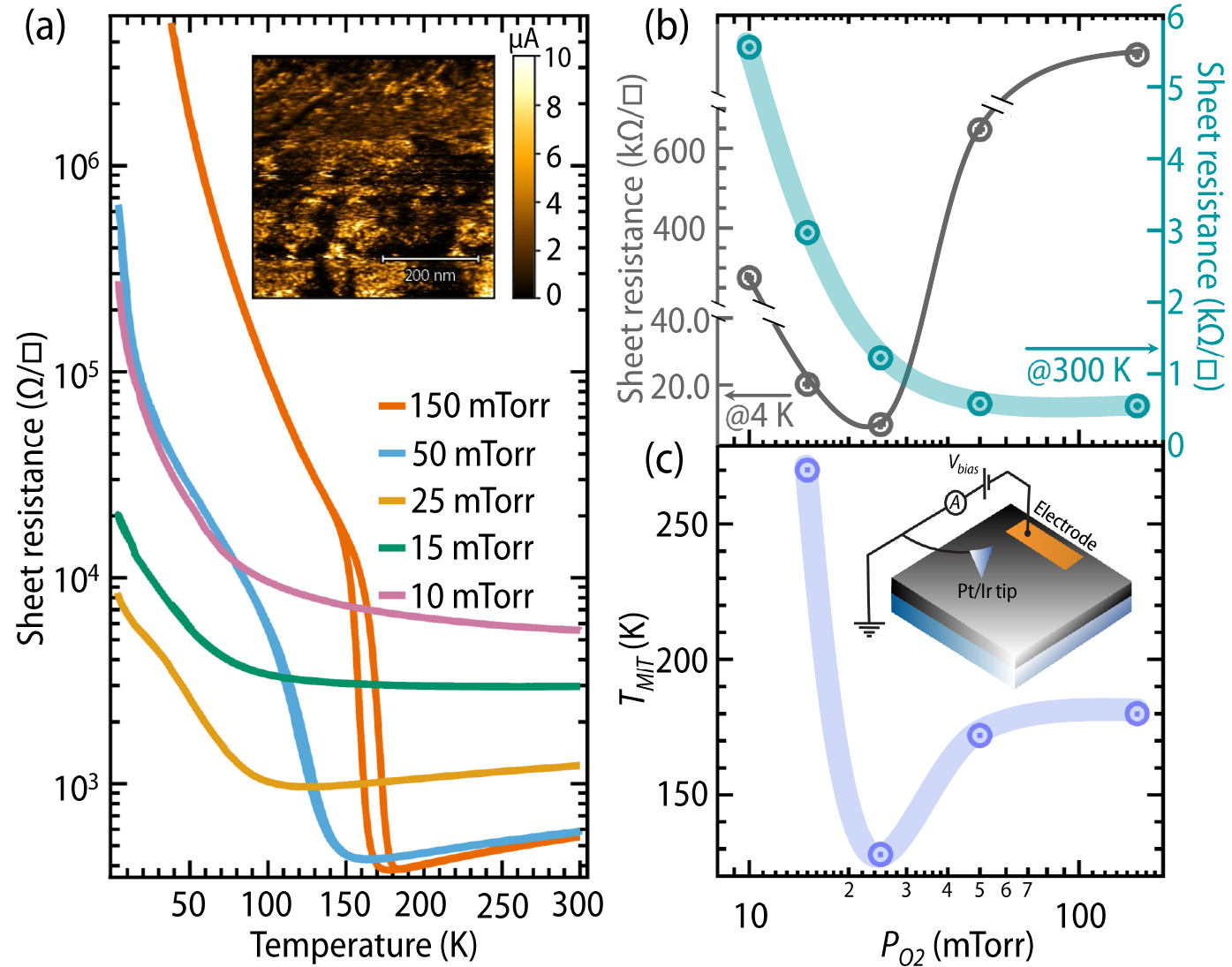}
    \caption{(a) Temperature-dependent sheet resistance of all the films.  (b) Room-temperature (300 K, right axis) and low-temperature (4 K, left axis) sheet resistance  and (c) MIT temperature ($T_\mathrm{MIT}$)  as a function of $P{_\mathrm{O2}}$.  The schematic setup for c-AFM is presented in the inset of panel (c). Room-temperature C-AFM imaging of the 25 mTorr sample, measured with a 3 V tip bias, is displayed in the inset of panel (a).}
    \label{Fig:3}
\end{figure}


The strong correlation between the electronic transport of (LPNSE)NO$_{3-\delta}$ films and the $P_{O_2}$	
  is evident in the sheet resistance ($R_s$) vs. temperature plot [Fig. \ref{Fig:3}(a)], measured using four probe Van der Pauw geometry.
We categorize the electronic phase as metallic (insulating) when $dR_S/dT$ is positive (negative) and define $T_\mathrm{MIT}$ as the temperature where $dR_S/dT$ becomes zero. The sample grown at 150 mTorr exhibits a first order MIT around 185 K with a strong hysteresis between the heating and cooling run. Interestingly, $T_\mathrm{MIT}$ decreases initially with the lowering of $P_{O_2}$. 
 The width of hysteresis also diminishes,  eventually disappearing entirely for the 25 mTorr sample. 
 For this sample, conductive-AFM imaging [Fig. \ref{Fig:3}(c) inset] further reveals room-temperature conduction inhomogeneity [inset of Fig. \ref{Fig:3}(a)]. Thus, the suppression of hysteresis width can be attributed to the impact of quenched disorder due to the increasing content of OVs~\cite{Guo:2020p2949}. With further lowering of $P_{O_2}$ to 15 mTorr, the  $T_\mathrm{MIT}$ increases to 285 K and the 10 mTorr sample is fully insulating below 300 K. As anticipated from the increased Ni$^{2+}$ content~\cite{Sanchez:1996p16574}, room-temperature $R_s$ exhibits a monotonic increase with decreasing $P_{O_2}$ [right axis of Fig.\ref{Fig:3}(b)].  However, a striking observation is the non-monotonic dependence of both the $T_\mathrm{MIT}$ and the 4 K sheet resistance on $P_{O_2}$ [Fig.\ref{Fig:3}(b), (c)]. This behavior sharply contrasts with recent studies on SmNiO$_{3-\delta}$, NdNiO$_{3-\delta}$, where  $T_\mathrm{MIT}$, $R_s$ at low temperature phase monotonically increase  with the increase of $\delta$~\cite{Kotiuga:2019p21992,Guo:2021p72}. For the 25 mTorr, 15 mTorr, and 10 mTorr samples,  the change in resistance between 300 K and 4 K is also significantly smaller [Fig.\ref{Fig:3}(a)], indicating that the nature of the insulating phase is strongly influenced by the oxygen pressure during growth and beyond previously reported carrier localization effect~\cite{Kotiuga:2019p21992}.

The critical oxygen partial pressure $P_{O_{2}}$ for the enhanced $T_\mathrm{MIT}$ in (LPNSE)NO$_{3-\delta}$ is 15 mTorr. The initial decrease in $T_\mathrm{MIT}$ with increasing OV concentration resembles the effect of electron doping via Ce$^{4+}$,Th$^{4+}$, and Pb$^{4+}$ substitution at the rare-earth ($RE$) site~\cite{Iglesias:2021p035123,Garc:1995p1356313569,Jiarui:2022p2296424X,Hadjimichael:2023p2201182}. In this regime, the lowering of $\Delta$ is the dominant factor as found from our O $K$-edge XAS~\cite{Middey:2018p045115,Liu:2013p2714}. Interestingly, the resistivity of the metallic phase shows linear dependence with temperature, sharply contrasting with conventional metallic behavior~\cite{Jaramillo:2014p304307}. 
Such linear resistivity has also been observed in a variety of compounds, including high $T_c$ superconductors, ruthenates, heavy fermions, etc~\cite{Bruin:2013p804}.
Remarkably, the scattering rate per Kelvin ($\sim k_B/\hbar$) is same in these systems, despite their differing microscopic scattering origins.

\begin{figure}[]
    \centering
    \includegraphics[width= .9\textwidth]{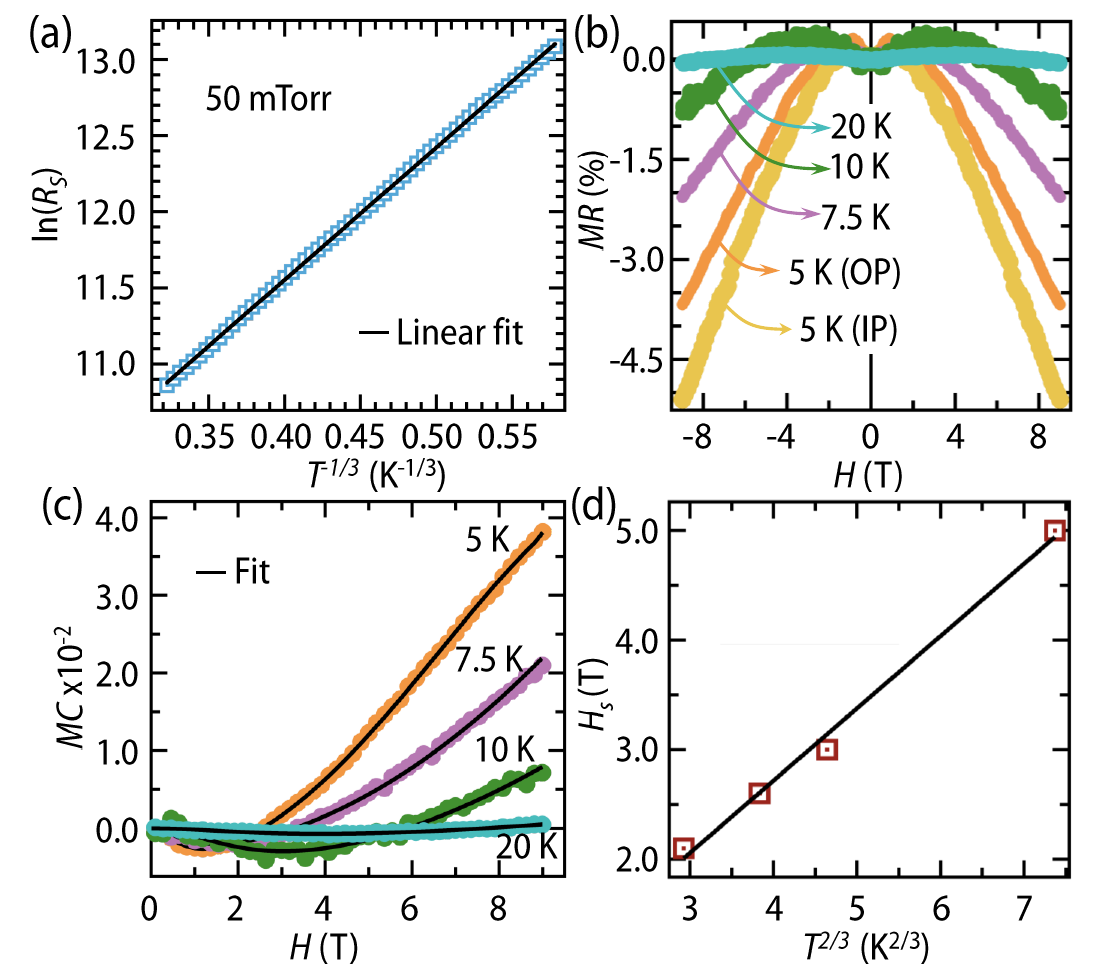}
    \caption{(a) Logarithm of sheet resistance as a function of $T^{-1/3}$ for 50 mTorr sample in the range of 4-30 K, illustrating the VRH regime in 2-dimension. (b) MR as a function of $H$, for both in-plane (IP) and out-of-plane (OP) field orientations relative to the sample.  (c) Fitting of the magneto conductance using the model described in main text. (d) $H_{s}$ as a function of $T^{2/3}$ showing linear behavior as described in the text.}
    \label{Fig:4}
\end{figure}

We now examine the impact of electron doping and disorder on the insulating phase by analyzing the temperature-dependent sheet resistance and low-temperature magnetotransport data. As the 150 mTorr sample exhibits a high $R_s$  below 50 K exceeding our measurement capabilities, we focus our analysis on the 50 mTorr sample. The low-temperature ($<$ 30 K) $R_s$ can be described by the formulae of Mott variable-range hopping (VRH) model in 2 dimensions [$R_s=R_0 \exp \left(\frac{T_{M}}{T}\right)^{1/3}$, see Fig.\ref{Fig:4}(a)] , where
 electron prefers to hop to a state that is closer in energy than the nearest neighbor. 
 For 2D VRH,  $T_M$ and hopping length ($R_{hop}$) is given by   $T_M \approx \frac{3}{k_B N(0) \xi^2 d}, R_{hop} \approx \left(\frac{T_M}{T}\right)^{1/3}\xi$, where $N(0)$ is the DOS near Fermi energy $E_F$, $\xi$ is localization length, and $d$ is the thickness of the film~\cite{Faran:1988p5457,Mott:1993p54575465}. Our fitting finds  $T_M$ = 673 K, $\xi$ of 2.93 nm, and $R_{hop}$ of 16.17 nm. The condition $\frac{R_{hop}}{\xi} > 1$ testifies the validity of the Mott VRH process in 2D. 3D VRH is ruled out as it yields $\xi>d$.

\begin{figure}[]
    \centering
    \includegraphics[width=0.8\textwidth]{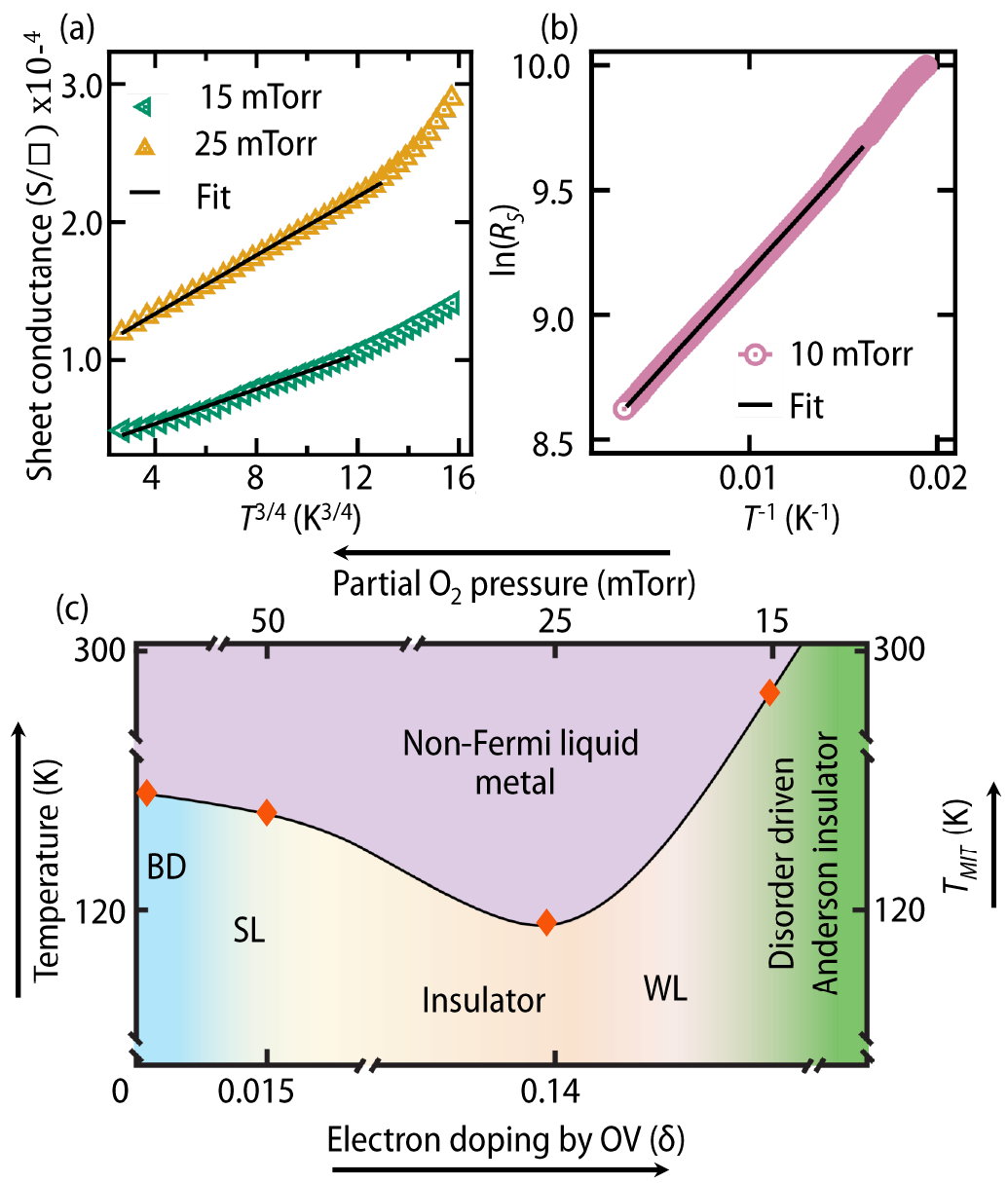}
    \caption{ (a) Sheet conductance as a function of T$^{3/4}$ for 25, and 15 mTorr samples. The black line is the linear fit of the data. (b) Logarithm of sheet resistance as a function of $T^{-1}$ for the 10 mTorr sample, the black line shows nearest neighbor hopping fit between 300 K and 60 K. (c) Phase diagram  as a function of $\delta$ and $P_{O2}$, showing realization of different  electronic phases in (LPNSE)NO$_{3-\delta}$.}
    \label{Fig:5}
\end{figure}

Fig. \ref{Fig:4}(b) displays the magnetoresistance (MR) of this sample in a perpendicular magnetic field (within $\pm$9 T).  The MR is positive at low fields and becomes negative at higher fields.  Furthermore, the field at which the MR switches from positive to negative decreases with decreasing temperature.
The positive MR of the VRH regime aligns with Frydman et al.'s model of exchange-correlation among spins in different hopping sites~\cite{Ramadoss:2016p235124,Frydman:1995p745}. According to this model, magnetoconductance (MC)         
 $  \frac{\Delta\sigma}{\sigma_0} = \frac{\sigma(H)-\sigma(0)}{\sigma(0)}=-A' \frac{H^2}{H^2 + H_{s}^2}$
where $A'$ is the saturation value, $a$ is constant and order of unity. The spin alignment field is
    $H_{s} = a\frac{k_BT}{\mu_B} (\frac{T_{M}}{T})^{1/3}$~\cite{Vaknin:1996p13604}.
The observed anisotropy in the negative MR with respect to the magnetic field direction (Fig. \ref{Fig:4}(b)) suggests an orbital, rather than spin-based, mechanism.  In the VRH regime, orbital quantum interference can lead to a positive MR that is proportional to $H^2$  in the low-field regime~\cite{Sivan:1988p1566,Faran:1988p5457}. 
Fig. \ref{Fig:4}(c) demonstrates excellent fits to the MR data at various temperatures using a combination of the aforementioned models.  Furthermore, the characteristic field $H_s$ obtained from these fits exhibits a $T^{2/3}$ 
(Fig. \ref{Fig:4}(d)), further validates our analysis.

$R_s$ of the films grown under 25 and 15 mTorr of $P_{O_2}$ are lower than the strong localization limit~\cite{Scherwitzl:2011p246403}.  Furthermore, the limited temperature range over which the VRH behavior is observed suggests that the insulating phases exhibit weak, rather than strong localization. The attempts to fit temperature dependence sheet conductance to a 2D weak localization (WL) model [$\sigma_{2d}(T) \sim ln(T)$] also failed. Furthermore, the 2D WL model can not explain our observed MC ~\cite{LeeRamakrishnan:1982p40094012,Scherwitzl:2011p246403}. A plausible scenario could be 3D WL with $ \sigma_{3d}(T) = \sigma_0 + \frac{e^2}{\hbar \pi^3 a} T^{p/2}$, where $\sigma_0$ is Drude conductivity, and $p$ is the temperature exponent of the inelastic scattering length $L_{in} \sim T^{-p}$. If the dominant scattering mechanism is due to electron-electron collisions, $p$ = 3/2, whereas electron-phonon scattering leads to $p$ = 2 ~\cite{BERGMANN:1984p158,Ramadoss:2016p235124,LeeRamakrishnan:1982p40094012}. As shown in Fig.~\ref{Fig:5}(a), the low-temperature transport behavior of both samples indeed aligns with 3D WL characterized by $p$=3/2. Magnetoconductance analysis of the 25 mTorr sample further confirms that the electron-electron scattering is the dominant mechanism within the 3D WL regime.

Finally, we focus on the film, grown under $P_{O_2}$ of 10 mTorr, exhibiting a fully insulating behavior. Unlike the site-selective Mott insulating behavior of LaNiO$_{3-\delta}$ ($\delta$=0.5) with its Ni$^{2+}$ states~\cite{Liao:2021p085110}, the predominantly Ni$^{3+}$ state in our film points to a different insulating mechanism. The temperature-dependence of $R_s$ between 300 K and 66 K  well-described by the nearest-neighbor hopping (NNH) conduction model [$R_s$=$R_0$exp($E_a$/$k_BT$)], which is observed very often in disordered systems~\cite{Guo:2020p2949,Guo:2021p72}. The thermal activation gap $E_a$ is found to be around 20 meV. 
Though distinguishing between 2D and 3D VRH is difficult at low temperatures, we primarily attribute this insulating phase to Anderson localization. To evaluate the role of $RE$-site disorder, we also grew and measured on the NdNiO$_{3-\delta}$ film under $P_{O_2}$=10 mTorr, following the same growth protocol. While insulating, this NdNiO$_{3-\delta}$ film exhibits lower resistivity and does not exhibit VRH behavior. This implies that the Mott-Anderson insulating phase of the HEO film is contributed by the disorder of the $RE$ sites as well as the oxygen sublattice.

In summary, using a model system (LPNSE)NO$_{3-\delta}$, we demonstrate that OVs serve as a powerful tunable parameter to engineer the electronic phase diagram within HEOs [Fig.~\ref{Fig:5}(c)].
By varying the $P_{O_2}$ during growth,  we experimentally access the low OV concentration region, where non-monotonic band gap dependence was previously predicted by density functional theory~\cite{Kotiuga:2019p21992} but experimentally inaccessible to prior studies. 
The insulating phase of the pristine, $\delta$ $\simeq$ $0$ film is a bond-disproportionate (BD) insulating state~\cite{Middey:2016p305334,Mazza:2023p013008}.  
Crucially, OVs act as potent disruptors of this BD order, likely by hindering polaron formation or condensation~\cite{Shamblin:2018p86}. This disruption directly manifests as an initial lowering of  $T_\mathrm {MIT}$ and the emergence of OV-induced local insulating domains embedded within a globally metallic matrix.  Pushing to higher OV concentrations, a fully realized Mott-Anderson insulating state is observed, due to the disorder within $RE$ and the oxygen sublattices, along with concomitant local crystal field symmetry lowering, and electron doping. Our work reveals a striking, Janus-faced response of $T_\mathrm {MIT}$ to OVs in this negative $\Delta$ system, extending far beyond the recently emphasized electron antidoping effect~\cite{Shi:2014p4860,Kotiuga:2019p21992,Li:2021p187602,Liu:2019p106403,Shi:2024p256502}. 
Given the indispensable role of carrier doping for transport-based device (electronics, optoelectronics, neuromorphic applications), this work provides a critical pathway for developing HEOs based electronic device technologies.

{\it Acknowledgement:} SJ and SM acknowledge Dr. Yin Shi, Dr. Long-Qing Chen, Dr. Venkatraman Gopalan, Dr. Shriram Ramanathan, and Dr. Nandini Trivedi for insightful discussions. The authors acknowledge the use of central facilities of the Department of Physics, IISc, funded through the FIST program of the Department of Science and Technology (DST), Gov. of India and the wire bonding facility of MEMS packaging lab, CENSE, IISc. SM  acknowledges funding support from a SERB Core Research grant (Grant No. CRG/2022/001906) and I.R.H.P.A Grant No. IPA/2020/000034.  NB and MB acknowledge funding from the Prime Minister’s Research Fellowship (PMRF), MoE, Government of India.
JM acknowledges UGC, India for fellowship. This research used resources of the Advanced Photon Source, a U.S. Department of Energy Office of Science User Facility operated by Argonne National Laboratory under Contract No. DE-AC02-06CH11357. This research used resources of the Advanced Light Source, which is a Department of Energy Office of Science User Facility under Contract No. DE-AC02-05CH11231.

\nocite{*}

\providecommand{\noopsort}[1]{}\providecommand{\singleletter}[1]{#1}%


\begin{thebibliography}{61}%
	\makeatletter
	\providecommand \@ifxundefined [1]{%
		\@ifx{#1\undefined}
	}%
	\providecommand \@ifnum [1]{%
		\ifnum #1\expandafter \@firstoftwo
		\else \expandafter \@secondoftwo
		\fi
	}%
	\providecommand \@ifx [1]{%
		\ifx #1\expandafter \@firstoftwo
		\else \expandafter \@secondoftwo
		\fi
	}%
	\providecommand \natexlab [1]{#1}%
	\providecommand \enquote  [1]{``#1''}%
	\providecommand \bibnamefont  [1]{#1}%
	\providecommand \bibfnamefont [1]{#1}%
	\providecommand \citenamefont [1]{#1}%
	\providecommand \href@noop [0]{\@secondoftwo}%
	\providecommand \href [0]{\begingroup \@sanitize@url \@href}%
	\providecommand \@href[1]{\@@startlink{#1}\@@href}%
	\providecommand \@@href[1]{\endgroup#1\@@endlink}%
	\providecommand \@sanitize@url [0]{\catcode `\\12\catcode `\$12\catcode
		`\&12\catcode `\#12\catcode `\^12\catcode `\_12\catcode `\%12\relax}%
	\providecommand \@@startlink[1]{}%
	\providecommand \@@endlink[0]{}%
	\providecommand \url  [0]{\begingroup\@sanitize@url \@url }%
	\providecommand \@url [1]{\endgroup\@href {#1}{\urlprefix }}%
	\providecommand \urlprefix  [0]{URL }%
	\providecommand \Eprint [0]{\href }%
	\providecommand \doibase [0]{https://doi.org/}%
	\providecommand \selectlanguage [0]{\@gobble}%
	\providecommand \bibinfo  [0]{\@secondoftwo}%
	\providecommand \bibfield  [0]{\@secondoftwo}%
	\providecommand \translation [1]{[#1]}%
	\providecommand \BibitemOpen [0]{}%
	\providecommand \bibitemStop [0]{}%
	\providecommand \bibitemNoStop [0]{.\EOS\space}%
	\providecommand \EOS [0]{\spacefactor3000\relax}%
	\providecommand \BibitemShut  [1]{\csname bibitem#1\endcsname}%
	\let\auto@bib@innerbib\@empty
	\bibitem [{\citenamefont {Yoo}\ \emph {et~al.}(2024)\citenamefont {Yoo},
		\citenamefont {Chae}, \citenamefont {Chiang}, \citenamefont {Webb},
		\citenamefont {Ma}, \citenamefont {Paik}, \citenamefont {Park}, \citenamefont
		{Williams}, \citenamefont {Nomoto}, \citenamefont {Xing}, \citenamefont
		{Trolier-McKinstry}, \citenamefont {Kioupakis}, \citenamefont {Heron},\ and\
		\citenamefont {Lu}}]{Yoo:2024p466474}%
	\BibitemOpen
	\bibfield  {author} {\bibinfo {author} {\bibfnamefont {S.}~\bibnamefont
			{Yoo}}, \bibinfo {author} {\bibfnamefont {S.}~\bibnamefont {Chae}}, \bibinfo
		{author} {\bibfnamefont {T.}~\bibnamefont {Chiang}}, \bibinfo {author}
		{\bibfnamefont {M.}~\bibnamefont {Webb}}, \bibinfo {author} {\bibfnamefont
			{T.}~\bibnamefont {Ma}}, \bibinfo {author} {\bibfnamefont {H.}~\bibnamefont
			{Paik}}, \bibinfo {author} {\bibfnamefont {Y.}~\bibnamefont {Park}}, \bibinfo
		{author} {\bibfnamefont {L.}~\bibnamefont {Williams}}, \bibinfo {author}
		{\bibfnamefont {K.}~\bibnamefont {Nomoto}}, \bibinfo {author} {\bibfnamefont
			{H.~G.}\ \bibnamefont {Xing}}, \bibinfo {author} {\bibfnamefont
			{S.}~\bibnamefont {Trolier-McKinstry}}, \bibinfo {author} {\bibfnamefont
			{E.}~\bibnamefont {Kioupakis}}, \bibinfo {author} {\bibfnamefont {J.~T.}\
			\bibnamefont {Heron}},\ and\ \bibinfo {author} {\bibfnamefont {W.~D.}\
			\bibnamefont {Lu}},\ }\bibfield  {title} {\bibinfo {title} {Efficient data
			processing using tunable entropy-stabilized oxide memristors},\ }\href
	{https://doi.org/10.1038/s41928-024-01169-1} {\bibfield  {journal} {\bibinfo
			{journal} {Nature Electronics}\ }\textbf {\bibinfo {volume} {7}},\ \bibinfo
		{pages} {466} (\bibinfo {year} {2024})}\BibitemShut {NoStop}%
	\bibitem [{\citenamefont {Tsai}\ \emph {et~al.}(2023)\citenamefont {Tsai},
		\citenamefont {Chen}, \citenamefont {Huang}, \citenamefont {Lo},
		\citenamefont {Ke}, \citenamefont {Chu},\ and\ \citenamefont
		{Wu}}]{Tsai:2023p2302979}%
	\BibitemOpen
	\bibfield  {author} {\bibinfo {author} {\bibfnamefont {J.-Y.}\ \bibnamefont
			{Tsai}}, \bibinfo {author} {\bibfnamefont {J.-Y.}\ \bibnamefont {Chen}},
		\bibinfo {author} {\bibfnamefont {C.-W.}\ \bibnamefont {Huang}}, \bibinfo
		{author} {\bibfnamefont {H.-Y.}\ \bibnamefont {Lo}}, \bibinfo {author}
		{\bibfnamefont {W.-E.}\ \bibnamefont {Ke}}, \bibinfo {author} {\bibfnamefont
			{Y.-H.}\ \bibnamefont {Chu}},\ and\ \bibinfo {author} {\bibfnamefont {W.-W.}\
			\bibnamefont {Wu}},\ }\bibfield  {title} {\bibinfo {title} {A
			high-entropy-oxides-based memristor: Outstanding resistive switching
			performance and mechanisms in atomic structural evolution},\ }\href
	{https://doi.org/https://doi.org/10.1002/adma.202302979} {\bibfield
		{journal} {\bibinfo  {journal} {Advanced Materials}\ }\textbf {\bibinfo
			{volume} {35}},\ \bibinfo {pages} {2302979} (\bibinfo {year}
		{2023})}\BibitemShut {NoStop}%
	\bibitem [{\citenamefont {Zou}\ \emph {et~al.}(2024)\citenamefont {Zou},
		\citenamefont {Tang}, \citenamefont {He},\ and\ \citenamefont
		{Zhang}}]{Zou:2024p3449234530}%
	\BibitemOpen
	\bibfield  {author} {\bibinfo {author} {\bibfnamefont {J.}~\bibnamefont
			{Zou}}, \bibinfo {author} {\bibfnamefont {L.}~\bibnamefont {Tang}}, \bibinfo
		{author} {\bibfnamefont {W.}~\bibnamefont {He}},\ and\ \bibinfo {author}
		{\bibfnamefont {X.}~\bibnamefont {Zhang}},\ }\bibfield  {title} {\bibinfo
		{title} {High-entropy oxides: Pioneering the future of multifunctional
			materials},\ }\href {https://doi.org/10.1021/acsnano.4c12538} {\bibfield
		{journal} {\bibinfo  {journal} {ACS Nano}\ }\textbf {\bibinfo {volume}
			{18}},\ \bibinfo {pages} {34492} (\bibinfo {year} {2024})}\BibitemShut
	{NoStop}%
	\bibitem [{\citenamefont {Oses}\ \emph {et~al.}(2020)\citenamefont {Oses},
		\citenamefont {Toher},\ and\ \citenamefont {Curtarolo}}]{Oses:2020P295309}%
	\BibitemOpen
	\bibfield  {author} {\bibinfo {author} {\bibfnamefont {C.}~\bibnamefont
			{Oses}}, \bibinfo {author} {\bibfnamefont {C.}~\bibnamefont {Toher}},\ and\
		\bibinfo {author} {\bibfnamefont {S.}~\bibnamefont {Curtarolo}},\ }\bibfield
	{title} {\bibinfo {title} {High-entropy ceramics},\ }\href
	{https://doi.org/10.1038/s41578-019-0170-8} {\bibfield  {journal} {\bibinfo
			{journal} {Nature Reviews Materials}\ }\textbf {\bibinfo {volume} {5}},\
		\bibinfo {pages} {295} (\bibinfo {year} {2020})}\BibitemShut {NoStop}%
	\bibitem [{\citenamefont {Sivak}\ \emph {et~al.}(2025)\citenamefont {Sivak},
		\citenamefont {Almishal}, \citenamefont {Caucci}, \citenamefont {Tan},
		\citenamefont {Srikanth}, \citenamefont {Petruska}, \citenamefont {Furst},
		\citenamefont {Chen}, \citenamefont {Rost}, \citenamefont {Maria},\ and\
		\citenamefont {Sinnott}}]{Sivak:2025p216101}%
	\BibitemOpen
	\bibfield  {author} {\bibinfo {author} {\bibfnamefont {J.~T.}\ \bibnamefont
			{Sivak}}, \bibinfo {author} {\bibfnamefont {S.~S.~I.}\ \bibnamefont
			{Almishal}}, \bibinfo {author} {\bibfnamefont {M.~K.}\ \bibnamefont
			{Caucci}}, \bibinfo {author} {\bibfnamefont {Y.}~\bibnamefont {Tan}},
		\bibinfo {author} {\bibfnamefont {D.}~\bibnamefont {Srikanth}}, \bibinfo
		{author} {\bibfnamefont {J.}~\bibnamefont {Petruska}}, \bibinfo {author}
		{\bibfnamefont {M.}~\bibnamefont {Furst}}, \bibinfo {author} {\bibfnamefont
			{L.-Q.}\ \bibnamefont {Chen}}, \bibinfo {author} {\bibfnamefont {C.~M.}\
			\bibnamefont {Rost}}, \bibinfo {author} {\bibfnamefont {J.-P.}\ \bibnamefont
			{Maria}},\ and\ \bibinfo {author} {\bibfnamefont {S.~B.}\ \bibnamefont
			{Sinnott}},\ }\bibfield  {title} {\bibinfo {title} {Discovering high-entropy
			oxides with a machine-learning interatomic potential},\ }\href
	{https://doi.org/10.1103/PhysRevLett.134.216101} {\bibfield  {journal}
		{\bibinfo  {journal} {Phys. Rev. Lett.}\ }\textbf {\bibinfo {volume} {134}},\
		\bibinfo {pages} {216101} (\bibinfo {year} {2025})}\BibitemShut {NoStop}%
	\bibitem [{\citenamefont {Mazza}\ \emph {et~al.}(2024)\citenamefont {Mazza},
		\citenamefont {Yan}, \citenamefont {Middey}, \citenamefont {Gardner},
		\citenamefont {Chen}, \citenamefont {Brahlek},\ and\ \citenamefont
		{Ward}}]{Mazza:2024p230501}%
	\BibitemOpen
	\bibfield  {author} {\bibinfo {author} {\bibfnamefont {A.~R.}\ \bibnamefont
			{Mazza}}, \bibinfo {author} {\bibfnamefont {J.-Q.}\ \bibnamefont {Yan}},
		\bibinfo {author} {\bibfnamefont {S.}~\bibnamefont {Middey}}, \bibinfo
		{author} {\bibfnamefont {J.~S.}\ \bibnamefont {Gardner}}, \bibinfo {author}
		{\bibfnamefont {A.-H.}\ \bibnamefont {Chen}}, \bibinfo {author}
		{\bibfnamefont {M.}~\bibnamefont {Brahlek}},\ and\ \bibinfo {author}
		{\bibfnamefont {T.~Z.}\ \bibnamefont {Ward}},\ }\bibfield  {title} {\bibinfo
		{title} {Embracing disorder in quantum materials design},\ }\href
	{https://doi.org/10.1063/5.0203647} {\bibfield  {journal} {\bibinfo
			{journal} {Applied Physics Letters}\ }\textbf {\bibinfo {volume} {124}},\
		\bibinfo {pages} {230501} (\bibinfo {year} {2024})}\BibitemShut {NoStop}%
	\bibitem [{\citenamefont {Xiong}\ \emph {et~al.}()\citenamefont {Xiong},
		\citenamefont {Liu}, \citenamefont {Zhang}, \citenamefont {Yang},
		\citenamefont {Liang}, \citenamefont {Zhou}, \citenamefont {Li},
		\citenamefont {Zhang}, \citenamefont {Lv},\ and\ \citenamefont
		{Che}}]{Xiong:2024p2415351}%
	\BibitemOpen
	\bibfield  {author} {\bibinfo {author} {\bibfnamefont {X.}~\bibnamefont
			{Xiong}}, \bibinfo {author} {\bibfnamefont {Z.}~\bibnamefont {Liu}}, \bibinfo
		{author} {\bibfnamefont {R.}~\bibnamefont {Zhang}}, \bibinfo {author}
		{\bibfnamefont {L.}~\bibnamefont {Yang}}, \bibinfo {author} {\bibfnamefont
			{G.}~\bibnamefont {Liang}}, \bibinfo {author} {\bibfnamefont
			{X.}~\bibnamefont {Zhou}}, \bibinfo {author} {\bibfnamefont {B.}~\bibnamefont
			{Li}}, \bibinfo {author} {\bibfnamefont {H.}~\bibnamefont {Zhang}}, \bibinfo
		{author} {\bibfnamefont {H.}~\bibnamefont {Lv}},\ and\ \bibinfo {author}
		{\bibfnamefont {R.}~\bibnamefont {Che}},\ }\bibfield  {title} {\bibinfo
		{title} {Atomic-level electric polarization in entropy-driven perovskites for
			boosting dielectric response},\ }\href
	{https://doi.org/https://doi.org/10.1002/adma.202415351} {\bibfield
		{journal} {\bibinfo  {journal} {Advanced Materials}\ }\textbf {\bibinfo
			{volume} {n/a}},\ \bibinfo {pages} {2415351}}\BibitemShut {NoStop}%
	\bibitem [{\citenamefont {Mazza}\ \emph {et~al.}(2022)\citenamefont {Mazza},
		\citenamefont {Skoropata}, \citenamefont {Sharma}, \citenamefont {Lapano},
		\citenamefont {Heitmann}, \citenamefont {Musico}, \citenamefont {Keppens},
		\citenamefont {Gai}, \citenamefont {Freeland}, \citenamefont {Charlton},
		\citenamefont {Brahlek}, \citenamefont {Moreo}, \citenamefont {Dagotto},\
		and\ \citenamefont {Ward}}]{Mazza:2022p2200391}%
	\BibitemOpen
	\bibfield  {author} {\bibinfo {author} {\bibfnamefont {A.~R.}\ \bibnamefont
			{Mazza}}, \bibinfo {author} {\bibfnamefont {E.}~\bibnamefont {Skoropata}},
		\bibinfo {author} {\bibfnamefont {Y.}~\bibnamefont {Sharma}}, \bibinfo
		{author} {\bibfnamefont {J.}~\bibnamefont {Lapano}}, \bibinfo {author}
		{\bibfnamefont {T.~W.}\ \bibnamefont {Heitmann}}, \bibinfo {author}
		{\bibfnamefont {B.~L.}\ \bibnamefont {Musico}}, \bibinfo {author}
		{\bibfnamefont {V.}~\bibnamefont {Keppens}}, \bibinfo {author} {\bibfnamefont
			{Z.}~\bibnamefont {Gai}}, \bibinfo {author} {\bibfnamefont {J.~W.}\
			\bibnamefont {Freeland}}, \bibinfo {author} {\bibfnamefont {T.~R.}\
			\bibnamefont {Charlton}}, \bibinfo {author} {\bibfnamefont {M.}~\bibnamefont
			{Brahlek}}, \bibinfo {author} {\bibfnamefont {A.}~\bibnamefont {Moreo}},
		\bibinfo {author} {\bibfnamefont {E.}~\bibnamefont {Dagotto}},\ and\ \bibinfo
		{author} {\bibfnamefont {T.~Z.}\ \bibnamefont {Ward}},\ }\bibfield  {title}
	{\bibinfo {title} {Designing magnetism in high entropy oxides},\ }\href
	{https://doi.org/https://doi.org/10.1002/advs.202200391} {\bibfield
		{journal} {\bibinfo  {journal} {Advanced Science}\ }\textbf {\bibinfo
			{volume} {9}},\ \bibinfo {pages} {2200391} (\bibinfo {year}
		{2022})}\BibitemShut {NoStop}%
	\bibitem [{\citenamefont {Zheng}\ \emph {et~al.}(2024)\citenamefont {Zheng},
		\citenamefont {Zhang}, \citenamefont {Shi}, \citenamefont {Zhou},
		\citenamefont {Lu}, \citenamefont {Han}, \citenamefont {Chen}, \citenamefont
		{Ma}, \citenamefont {Zhang}, \citenamefont {Lin}, \citenamefont {Xu},
		\citenamefont {Ma}, \citenamefont {Li}, \citenamefont {Yang}, \citenamefont
		{Wei}, \citenamefont {Yang}, \citenamefont {Zou}, \citenamefont {Zhang},
		\citenamefont {Liu}, \citenamefont {Dou}, \citenamefont {Yang}, \citenamefont
		{Lan}, \citenamefont {Yi}, \citenamefont {Zhang}, \citenamefont {Gu},
		\citenamefont {Nan},\ and\ \citenamefont {Lin}}]{Zheng:2024p7650}%
	\BibitemOpen
	\bibfield  {author} {\bibinfo {author} {\bibfnamefont {Y.}~\bibnamefont
			{Zheng}}, \bibinfo {author} {\bibfnamefont {Q.}~\bibnamefont {Zhang}},
		\bibinfo {author} {\bibfnamefont {C.}~\bibnamefont {Shi}}, \bibinfo {author}
		{\bibfnamefont {Z.}~\bibnamefont {Zhou}}, \bibinfo {author} {\bibfnamefont
			{Y.}~\bibnamefont {Lu}}, \bibinfo {author} {\bibfnamefont {J.}~\bibnamefont
			{Han}}, \bibinfo {author} {\bibfnamefont {H.}~\bibnamefont {Chen}}, \bibinfo
		{author} {\bibfnamefont {Y.}~\bibnamefont {Ma}}, \bibinfo {author}
		{\bibfnamefont {Y.}~\bibnamefont {Zhang}}, \bibinfo {author} {\bibfnamefont
			{C.}~\bibnamefont {Lin}}, \bibinfo {author} {\bibfnamefont {W.}~\bibnamefont
			{Xu}}, \bibinfo {author} {\bibfnamefont {W.}~\bibnamefont {Ma}}, \bibinfo
		{author} {\bibfnamefont {Q.}~\bibnamefont {Li}}, \bibinfo {author}
		{\bibfnamefont {Y.}~\bibnamefont {Yang}}, \bibinfo {author} {\bibfnamefont
			{B.}~\bibnamefont {Wei}}, \bibinfo {author} {\bibfnamefont {B.}~\bibnamefont
			{Yang}}, \bibinfo {author} {\bibfnamefont {M.}~\bibnamefont {Zou}}, \bibinfo
		{author} {\bibfnamefont {W.}~\bibnamefont {Zhang}}, \bibinfo {author}
		{\bibfnamefont {C.}~\bibnamefont {Liu}}, \bibinfo {author} {\bibfnamefont
			{L.}~\bibnamefont {Dou}}, \bibinfo {author} {\bibfnamefont {D.}~\bibnamefont
			{Yang}}, \bibinfo {author} {\bibfnamefont {J.-L.}\ \bibnamefont {Lan}},
		\bibinfo {author} {\bibfnamefont {D.}~\bibnamefont {Yi}}, \bibinfo {author}
		{\bibfnamefont {X.}~\bibnamefont {Zhang}}, \bibinfo {author} {\bibfnamefont
			{L.}~\bibnamefont {Gu}}, \bibinfo {author} {\bibfnamefont {C.-W.}\
			\bibnamefont {Nan}},\ and\ \bibinfo {author} {\bibfnamefont {Y.-H.}\
			\bibnamefont {Lin}},\ }\bibfield  {title} {\bibinfo {title} {Carrier-phonon
			decoupling in perovskite thermoelectrics via entropy engineering},\ }\href
	{https://doi.org/10.1038/s41467-024-52063-5} {\bibfield  {journal} {\bibinfo
			{journal} {Nature Communications}\ }\textbf {\bibinfo {volume} {15}},\
		\bibinfo {pages} {7650} (\bibinfo {year} {2024})}\BibitemShut {NoStop}%
	\bibitem [{\citenamefont {Zhang}\ \emph
		{et~al.}(2024{\natexlab{a}})\citenamefont {Zhang}, \citenamefont {Oh},
		\citenamefont {Choi}, \citenamefont {Rotenberg}, \citenamefont {Brown},
		\citenamefont {Spataru}, \citenamefont {Kinigstein}, \citenamefont {Guo},
		\citenamefont {Sugar}, \citenamefont {Salagre}, \citenamefont {Mascaraque},
		\citenamefont {Michel}, \citenamefont {Shad}, \citenamefont {Zhu},
		\citenamefont {Witman}, \citenamefont {Kumar}, \citenamefont {Talin},\ and\
		\citenamefont {Fuller}}]{Zhang:2024p2406885}%
	\BibitemOpen
	\bibfield  {author} {\bibinfo {author} {\bibfnamefont {A.}~\bibnamefont
			{Zhang}}, \bibinfo {author} {\bibfnamefont {S.}~\bibnamefont {Oh}}, \bibinfo
		{author} {\bibfnamefont {B.~K.}\ \bibnamefont {Choi}}, \bibinfo {author}
		{\bibfnamefont {E.}~\bibnamefont {Rotenberg}}, \bibinfo {author}
		{\bibfnamefont {T.~D.}\ \bibnamefont {Brown}}, \bibinfo {author}
		{\bibfnamefont {C.~D.}\ \bibnamefont {Spataru}}, \bibinfo {author}
		{\bibfnamefont {E.}~\bibnamefont {Kinigstein}}, \bibinfo {author}
		{\bibfnamefont {J.}~\bibnamefont {Guo}}, \bibinfo {author} {\bibfnamefont
			{J.~D.}\ \bibnamefont {Sugar}}, \bibinfo {author} {\bibfnamefont
			{E.}~\bibnamefont {Salagre}}, \bibinfo {author} {\bibfnamefont
			{A.}~\bibnamefont {Mascaraque}}, \bibinfo {author} {\bibfnamefont {E.~G.}\
			\bibnamefont {Michel}}, \bibinfo {author} {\bibfnamefont {A.~C.}\
			\bibnamefont {Shad}}, \bibinfo {author} {\bibfnamefont {J.}~\bibnamefont
			{Zhu}}, \bibinfo {author} {\bibfnamefont {M.~D.}\ \bibnamefont {Witman}},
		\bibinfo {author} {\bibfnamefont {S.}~\bibnamefont {Kumar}}, \bibinfo
		{author} {\bibfnamefont {A.~A.}\ \bibnamefont {Talin}},\ and\ \bibinfo
		{author} {\bibfnamefont {E.~J.}\ \bibnamefont {Fuller}},\ }\bibfield  {title}
	{\bibinfo {title} {Tuning the spin transition and carrier type in rare-earth
			cobaltates via compositional complexity},\ }\href
	{https://doi.org/10.1002/adma.202406885} {\bibfield  {journal} {\bibinfo
			{journal} {Advanced Materials}\ }\textbf {\bibinfo {volume} {36}},\ \bibinfo
		{pages} {2406885} (\bibinfo {year} {2024}{\natexlab{a}})}\BibitemShut
	{NoStop}%
	\bibitem [{\citenamefont {Zhang}\ \emph
		{et~al.}(2024{\natexlab{b}})\citenamefont {Zhang}, \citenamefont {Lan},
		\citenamefont {Yang}, \citenamefont {Pan}, \citenamefont {Liu}, \citenamefont
		{Zhang}, \citenamefont {Qi}, \citenamefont {Chen}, \citenamefont {Su},
		\citenamefont {Yi}, \citenamefont {Yang}, \citenamefont {Wei}, \citenamefont
		{Cai}, \citenamefont {Han}, \citenamefont {Gu}, \citenamefont {Nan},\ and\
		\citenamefont {Lin}}]{Zhang:2024p185189}%
	\BibitemOpen
	\bibfield  {author} {\bibinfo {author} {\bibfnamefont {M.}~\bibnamefont
			{Zhang}}, \bibinfo {author} {\bibfnamefont {S.}~\bibnamefont {Lan}}, \bibinfo
		{author} {\bibfnamefont {B.~B.}\ \bibnamefont {Yang}}, \bibinfo {author}
		{\bibfnamefont {H.}~\bibnamefont {Pan}}, \bibinfo {author} {\bibfnamefont
			{Y.~Q.}\ \bibnamefont {Liu}}, \bibinfo {author} {\bibfnamefont {Q.~H.}\
			\bibnamefont {Zhang}}, \bibinfo {author} {\bibfnamefont {J.~L.}\ \bibnamefont
			{Qi}}, \bibinfo {author} {\bibfnamefont {D.}~\bibnamefont {Chen}}, \bibinfo
		{author} {\bibfnamefont {H.}~\bibnamefont {Su}}, \bibinfo {author}
		{\bibfnamefont {D.}~\bibnamefont {Yi}}, \bibinfo {author} {\bibfnamefont
			{Y.~Y.}\ \bibnamefont {Yang}}, \bibinfo {author} {\bibfnamefont
			{R.}~\bibnamefont {Wei}}, \bibinfo {author} {\bibfnamefont {H.~D.}\
			\bibnamefont {Cai}}, \bibinfo {author} {\bibfnamefont {H.~J.}\ \bibnamefont
			{Han}}, \bibinfo {author} {\bibfnamefont {L.}~\bibnamefont {Gu}}, \bibinfo
		{author} {\bibfnamefont {C.-W.}\ \bibnamefont {Nan}},\ and\ \bibinfo {author}
		{\bibfnamefont {Y.-H.}\ \bibnamefont {Lin}},\ }\bibfield  {title} {\bibinfo
		{title} {Ultrahigh energy storage in high-entropy ceramic capacitors with
			polymorphic relaxor phase},\ }\href {https://doi.org/10.1126/science.adl2931}
	{\bibfield  {journal} {\bibinfo  {journal} {Science}\ }\textbf {\bibinfo
			{volume} {384}},\ \bibinfo {pages} {185} (\bibinfo {year}
		{2024}{\natexlab{b}})}\BibitemShut {NoStop}%
	\bibitem [{\citenamefont {Yang}\ \emph {et~al.}(2022)\citenamefont {Yang},
		\citenamefont {Zhang}, \citenamefont {Pan}, \citenamefont {Si}, \citenamefont
		{Zhang}, \citenamefont {Shen}, \citenamefont {Yu}, \citenamefont {Lan},
		\citenamefont {Meng}, \citenamefont {Liu}, \citenamefont {Huang},
		\citenamefont {He}, \citenamefont {Gu}, \citenamefont {Zhang}, \citenamefont
		{Chen}, \citenamefont {Zhu}, \citenamefont {Nan},\ and\ \citenamefont
		{Lin}}]{Yang:2022p1074-1080}%
	\BibitemOpen
	\bibfield  {author} {\bibinfo {author} {\bibfnamefont {B.}~\bibnamefont
			{Yang}}, \bibinfo {author} {\bibfnamefont {Y.}~\bibnamefont {Zhang}},
		\bibinfo {author} {\bibfnamefont {H.}~\bibnamefont {Pan}}, \bibinfo {author}
		{\bibfnamefont {W.}~\bibnamefont {Si}}, \bibinfo {author} {\bibfnamefont
			{Q.}~\bibnamefont {Zhang}}, \bibinfo {author} {\bibfnamefont
			{Z.}~\bibnamefont {Shen}}, \bibinfo {author} {\bibfnamefont {Y.}~\bibnamefont
			{Yu}}, \bibinfo {author} {\bibfnamefont {S.}~\bibnamefont {Lan}}, \bibinfo
		{author} {\bibfnamefont {F.}~\bibnamefont {Meng}}, \bibinfo {author}
		{\bibfnamefont {Y.}~\bibnamefont {Liu}}, \bibinfo {author} {\bibfnamefont
			{H.}~\bibnamefont {Huang}}, \bibinfo {author} {\bibfnamefont
			{J.}~\bibnamefont {He}}, \bibinfo {author} {\bibfnamefont {L.}~\bibnamefont
			{Gu}}, \bibinfo {author} {\bibfnamefont {S.}~\bibnamefont {Zhang}}, \bibinfo
		{author} {\bibfnamefont {L.-Q.}\ \bibnamefont {Chen}}, \bibinfo {author}
		{\bibfnamefont {J.}~\bibnamefont {Zhu}}, \bibinfo {author} {\bibfnamefont
			{C.-W.}\ \bibnamefont {Nan}},\ and\ \bibinfo {author} {\bibfnamefont {Y.-H.}\
			\bibnamefont {Lin}},\ }\bibfield  {title} {\bibinfo {title} {High-entropy
			enhanced capacitive energy storage},\ }\href
	{https://doi.org/10.1038/s41563-022-01274-6} {\bibfield  {journal} {\bibinfo
			{journal} {Nature Materials}\ }\textbf {\bibinfo {volume} {21}},\ \bibinfo
		{pages} {1074} (\bibinfo {year} {2022})}\BibitemShut {NoStop}%
	\bibitem [{\citenamefont {Sarkar}\ \emph {et~al.}(2019)\citenamefont {Sarkar},
		\citenamefont {Wang}, \citenamefont {Schiele}, \citenamefont {Chellali},
		\citenamefont {Bhattacharya}, \citenamefont {Wang}, \citenamefont
		{Brezesinski}, \citenamefont {Hahn}, \citenamefont {Velasco},\ and\
		\citenamefont {Breitung}}]{Sarkar:2019p1806236}%
	\BibitemOpen
	\bibfield  {author} {\bibinfo {author} {\bibfnamefont {A.}~\bibnamefont
			{Sarkar}}, \bibinfo {author} {\bibfnamefont {Q.}~\bibnamefont {Wang}},
		\bibinfo {author} {\bibfnamefont {A.}~\bibnamefont {Schiele}}, \bibinfo
		{author} {\bibfnamefont {M.~R.}\ \bibnamefont {Chellali}}, \bibinfo {author}
		{\bibfnamefont {S.~S.}\ \bibnamefont {Bhattacharya}}, \bibinfo {author}
		{\bibfnamefont {D.}~\bibnamefont {Wang}}, \bibinfo {author} {\bibfnamefont
			{T.}~\bibnamefont {Brezesinski}}, \bibinfo {author} {\bibfnamefont
			{H.}~\bibnamefont {Hahn}}, \bibinfo {author} {\bibfnamefont {L.}~\bibnamefont
			{Velasco}},\ and\ \bibinfo {author} {\bibfnamefont {B.}~\bibnamefont
			{Breitung}},\ }\bibfield  {title} {\bibinfo {title} {High-entropy oxides:
			Fundamental aspects and electrochemical properties},\ }\href
	{https://doi.org/https://doi.org/10.1002/adma.201806236} {\bibfield
		{journal} {\bibinfo  {journal} {Advanced Materials}\ }\textbf {\bibinfo
			{volume} {31}},\ \bibinfo {pages} {1806236} (\bibinfo {year}
		{2019})}\BibitemShut {NoStop}%
	\bibitem [{\citenamefont {Patel}\ \emph {et~al.}(2023)\citenamefont {Patel},
		\citenamefont {Jenjeti}, \citenamefont {Kumar}, \citenamefont {Bhattacharya},
		\citenamefont {Kumar}, \citenamefont {Ojha}, \citenamefont {Zhang},
		\citenamefont {Zhou}, \citenamefont {Qu}, \citenamefont {Wang}, \citenamefont
		{Yang}, \citenamefont {Klewe}, \citenamefont {Shafer}, \citenamefont
		{Sampath},\ and\ \citenamefont {Middey}}]{Patel:2023p031407}%
	\BibitemOpen
	\bibfield  {author} {\bibinfo {author} {\bibfnamefont {R.~K.}\ \bibnamefont
			{Patel}}, \bibinfo {author} {\bibfnamefont {R.~N.}\ \bibnamefont {Jenjeti}},
		\bibinfo {author} {\bibfnamefont {R.}~\bibnamefont {Kumar}}, \bibinfo
		{author} {\bibfnamefont {N.}~\bibnamefont {Bhattacharya}}, \bibinfo {author}
		{\bibfnamefont {S.}~\bibnamefont {Kumar}}, \bibinfo {author} {\bibfnamefont
			{S.~K.}\ \bibnamefont {Ojha}}, \bibinfo {author} {\bibfnamefont
			{Z.}~\bibnamefont {Zhang}}, \bibinfo {author} {\bibfnamefont
			{H.}~\bibnamefont {Zhou}}, \bibinfo {author} {\bibfnamefont {K.}~\bibnamefont
			{Qu}}, \bibinfo {author} {\bibfnamefont {Z.}~\bibnamefont {Wang}}, \bibinfo
		{author} {\bibfnamefont {Z.}~\bibnamefont {Yang}}, \bibinfo {author}
		{\bibfnamefont {C.}~\bibnamefont {Klewe}}, \bibinfo {author} {\bibfnamefont
			{P.}~\bibnamefont {Shafer}}, \bibinfo {author} {\bibfnamefont
			{S.}~\bibnamefont {Sampath}},\ and\ \bibinfo {author} {\bibfnamefont
			{S.}~\bibnamefont {Middey}},\ }\bibfield  {title} {\bibinfo {title}
		{{Thickness dependent OER electrocatalysis of epitaxial thin film of high
				entropy oxide}},\ }\href {https://doi.org/10.1063/5.0146005} {\bibfield
		{journal} {\bibinfo  {journal} {Applied Physics Reviews}\ }\textbf {\bibinfo
			{volume} {10}},\ \bibinfo {pages} {031407} (\bibinfo {year}
		{2023})}\BibitemShut {NoStop}%
	\bibitem [{\citenamefont {Anderson}(1958)}]{Anderson:1958p1492}%
	\BibitemOpen
	\bibfield  {author} {\bibinfo {author} {\bibfnamefont {P.~W.}\ \bibnamefont
			{Anderson}},\ }\bibfield  {title} {\bibinfo {title} {Absence of diffusion in
			certain random lattices},\ }\href@noop {} {\bibfield  {journal} {\bibinfo
			{journal} {Physical review}\ }\textbf {\bibinfo {volume} {109}},\ \bibinfo
		{pages} {1492} (\bibinfo {year} {1958})}\BibitemShut {NoStop}%
	\bibitem [{\citenamefont {Amir}\ \emph {et~al.}(2011)\citenamefont {Amir},
		\citenamefont {Oreg},\ and\ \citenamefont {Imry}}]{Amir:2011p235}%
	\BibitemOpen
	\bibfield  {author} {\bibinfo {author} {\bibfnamefont {A.}~\bibnamefont
			{Amir}}, \bibinfo {author} {\bibfnamefont {Y.}~\bibnamefont {Oreg}},\ and\
		\bibinfo {author} {\bibfnamefont {Y.}~\bibnamefont {Imry}},\ }\bibfield
	{title} {\bibinfo {title} {Electron glass dynamics},\ }\href@noop {}
	{\bibfield  {journal} {\bibinfo  {journal} {Annu. Rev. Condens. Matter
				Phys.}\ }\textbf {\bibinfo {volume} {2}},\ \bibinfo {pages} {235} (\bibinfo
		{year} {2011})}\BibitemShut {NoStop}%
	\bibitem [{\citenamefont {Abanin}\ \emph {et~al.}(2019)\citenamefont {Abanin},
		\citenamefont {Altman}, \citenamefont {Bloch},\ and\ \citenamefont
		{Serbyn}}]{Abanin:2019p021001}%
	\BibitemOpen
	\bibfield  {author} {\bibinfo {author} {\bibfnamefont {D.~A.}\ \bibnamefont
			{Abanin}}, \bibinfo {author} {\bibfnamefont {E.}~\bibnamefont {Altman}},
		\bibinfo {author} {\bibfnamefont {I.}~\bibnamefont {Bloch}},\ and\ \bibinfo
		{author} {\bibfnamefont {M.}~\bibnamefont {Serbyn}},\ }\bibfield  {title}
	{\bibinfo {title} {Colloquium: Many-body localization, thermalization, and
			entanglement},\ }\href {https://doi.org/10.1103/RevModPhys.91.021001}
	{\bibfield  {journal} {\bibinfo  {journal} {Rev. Mod. Phys.}\ }\textbf
		{\bibinfo {volume} {91}},\ \bibinfo {pages} {021001} (\bibinfo {year}
		{2019})}\BibitemShut {NoStop}%
	\bibitem [{\citenamefont {Patel}\ \emph {et~al.}(2020)\citenamefont {Patel},
		\citenamefont {Ojha}, \citenamefont {Kumar}, \citenamefont {Saha},
		\citenamefont {Mandal}, \citenamefont {Freeland},\ and\ \citenamefont
		{Middey}}]{Patel:2020p071601}%
	\BibitemOpen
	\bibfield  {author} {\bibinfo {author} {\bibfnamefont {R.~K.}\ \bibnamefont
			{Patel}}, \bibinfo {author} {\bibfnamefont {S.~K.}\ \bibnamefont {Ojha}},
		\bibinfo {author} {\bibfnamefont {S.}~\bibnamefont {Kumar}}, \bibinfo
		{author} {\bibfnamefont {A.}~\bibnamefont {Saha}}, \bibinfo {author}
		{\bibfnamefont {P.}~\bibnamefont {Mandal}}, \bibinfo {author} {\bibfnamefont
			{J.~W.}\ \bibnamefont {Freeland}},\ and\ \bibinfo {author} {\bibfnamefont
			{S.}~\bibnamefont {Middey}},\ }\bibfield  {title} {\bibinfo {title}
		{{Epitaxial stabilization of ultra thin films of high entropy perovskite}},\
	}\href {https://doi.org/10.1063/1.5133710} {\bibfield  {journal} {\bibinfo
			{journal} {Applied Physics Letters}\ }\textbf {\bibinfo {volume} {116}},\
		\bibinfo {pages} {071601} (\bibinfo {year} {2020})}\BibitemShut {NoStop}%
	\bibitem [{\citenamefont {Middey}\ \emph {et~al.}(2016)\citenamefont {Middey},
		\citenamefont {Chakhalian}, \citenamefont {Mahadevan}, \citenamefont
		{Freeland}, \citenamefont {Millis},\ and\ \citenamefont
		{Sarma}}]{Middey:2016p305334}%
	\BibitemOpen
	\bibfield  {author} {\bibinfo {author} {\bibfnamefont {S.}~\bibnamefont
			{Middey}}, \bibinfo {author} {\bibfnamefont {J.}~\bibnamefont {Chakhalian}},
		\bibinfo {author} {\bibfnamefont {P.}~\bibnamefont {Mahadevan}}, \bibinfo
		{author} {\bibfnamefont {J.}~\bibnamefont {Freeland}}, \bibinfo {author}
		{\bibfnamefont {A.}~\bibnamefont {Millis}},\ and\ \bibinfo {author}
		{\bibfnamefont {D.}~\bibnamefont {Sarma}},\ }\bibfield  {title} {\bibinfo
		{title} {Physics of ultrathin films and heterostructures of rare-earth
			nickelates},\ }\href
	{https://doi.org/https://doi.org/10.1146/annurev-matsci-070115-032057}
	{\bibfield  {journal} {\bibinfo  {journal} {Annual Review of Materials
				Research}\ }\textbf {\bibinfo {volume} {46}},\ \bibinfo {pages} {305}
		(\bibinfo {year} {2016})}\BibitemShut {NoStop}%
	\bibitem [{\citenamefont {Park}\ \emph {et~al.}(2012)\citenamefont {Park},
		\citenamefont {Millis},\ and\ \citenamefont {Marianetti}}]{Park:2012p156402}%
	\BibitemOpen
	\bibfield  {author} {\bibinfo {author} {\bibfnamefont {H.}~\bibnamefont
			{Park}}, \bibinfo {author} {\bibfnamefont {A.~J.}\ \bibnamefont {Millis}},\
		and\ \bibinfo {author} {\bibfnamefont {C.~A.}\ \bibnamefont {Marianetti}},\
	}\bibfield  {title} {\bibinfo {title} {Site-selective mott transition in
			rare-earth-element nickelates},\ }\href
	{https://doi.org/10.1103/PhysRevLett.109.156402} {\bibfield  {journal}
		{\bibinfo  {journal} {Phys. Rev. Lett.}\ }\textbf {\bibinfo {volume} {109}},\
		\bibinfo {pages} {156402} (\bibinfo {year} {2012})}\BibitemShut {NoStop}%
	\bibitem [{\citenamefont {Bisogni}\ \emph {et~al.}(2016)\citenamefont
		{Bisogni}, \citenamefont {Catalano}, \citenamefont {Green}, \citenamefont
		{Gibert}, \citenamefont {Scherwitzl}, \citenamefont {Huang}, \citenamefont
		{Strocov}, \citenamefont {Zubko}, \citenamefont {Balandeh}, \citenamefont
		{Triscone} \emph {et~al.}}]{Bisogni:2016p13017}%
	\BibitemOpen
	\bibfield  {author} {\bibinfo {author} {\bibfnamefont {V.}~\bibnamefont
			{Bisogni}}, \bibinfo {author} {\bibfnamefont {S.}~\bibnamefont {Catalano}},
		\bibinfo {author} {\bibfnamefont {R.~J.}\ \bibnamefont {Green}}, \bibinfo
		{author} {\bibfnamefont {M.}~\bibnamefont {Gibert}}, \bibinfo {author}
		{\bibfnamefont {R.}~\bibnamefont {Scherwitzl}}, \bibinfo {author}
		{\bibfnamefont {Y.}~\bibnamefont {Huang}}, \bibinfo {author} {\bibfnamefont
			{V.~N.}\ \bibnamefont {Strocov}}, \bibinfo {author} {\bibfnamefont
			{P.}~\bibnamefont {Zubko}}, \bibinfo {author} {\bibfnamefont
			{S.}~\bibnamefont {Balandeh}}, \bibinfo {author} {\bibfnamefont {J.-M.}\
			\bibnamefont {Triscone}}, \emph {et~al.},\ }\bibfield  {title} {\bibinfo
		{title} {Ground-state oxygen holes and the metal--insulator transition in the
			negative charge-transfer rare-earth nickelates},\ }\href@noop {} {\bibfield
		{journal} {\bibinfo  {journal} {Nature Communications}\ }\textbf {\bibinfo
			{volume} {7}},\ \bibinfo {pages} {13017} (\bibinfo {year}
		{2016})}\BibitemShut {NoStop}%
	\bibitem [{\citenamefont {Middey}\ \emph
		{et~al.}(2018{\natexlab{a}})\citenamefont {Middey}, \citenamefont {Meyers},
		\citenamefont {Kareev}, \citenamefont {Cao}, \citenamefont {Liu},
		\citenamefont {Shafer}, \citenamefont {Freeland}, \citenamefont {Kim},
		\citenamefont {Ryan},\ and\ \citenamefont {Chakhalian}}]{Middey:2018p156801}%
	\BibitemOpen
	\bibfield  {author} {\bibinfo {author} {\bibfnamefont {S.}~\bibnamefont
			{Middey}}, \bibinfo {author} {\bibfnamefont {D.}~\bibnamefont {Meyers}},
		\bibinfo {author} {\bibfnamefont {M.}~\bibnamefont {Kareev}}, \bibinfo
		{author} {\bibfnamefont {Y.}~\bibnamefont {Cao}}, \bibinfo {author}
		{\bibfnamefont {X.}~\bibnamefont {Liu}}, \bibinfo {author} {\bibfnamefont
			{P.}~\bibnamefont {Shafer}}, \bibinfo {author} {\bibfnamefont {J.~W.}\
			\bibnamefont {Freeland}}, \bibinfo {author} {\bibfnamefont {J.-W.}\
			\bibnamefont {Kim}}, \bibinfo {author} {\bibfnamefont {P.~J.}\ \bibnamefont
			{Ryan}},\ and\ \bibinfo {author} {\bibfnamefont {J.}~\bibnamefont
			{Chakhalian}},\ }\bibfield  {title} {\bibinfo {title} {Disentangled
			cooperative orderings in artificial rare-earth nickelates},\ }\href
	{https://doi.org/10.1103/PhysRevLett.120.156801} {\bibfield  {journal}
		{\bibinfo  {journal} {Phys. Rev. Lett.}\ }\textbf {\bibinfo {volume} {120}},\
		\bibinfo {pages} {156801} (\bibinfo {year} {2018}{\natexlab{a}})}\BibitemShut
	{NoStop}%
	\bibitem [{\citenamefont {Shamblin}\ \emph {et~al.}(2018)\citenamefont
		{Shamblin}, \citenamefont {Heres}, \citenamefont {Zhou}, \citenamefont
		{Sangoro}, \citenamefont {Lang}, \citenamefont {Neuefeind}, \citenamefont
		{Alonso},\ and\ \citenamefont {Johnston}}]{Shamblin:2018p86}%
	\BibitemOpen
	\bibfield  {author} {\bibinfo {author} {\bibfnamefont {J.}~\bibnamefont
			{Shamblin}}, \bibinfo {author} {\bibfnamefont {M.}~\bibnamefont {Heres}},
		\bibinfo {author} {\bibfnamefont {H.}~\bibnamefont {Zhou}}, \bibinfo {author}
		{\bibfnamefont {J.}~\bibnamefont {Sangoro}}, \bibinfo {author} {\bibfnamefont
			{M.}~\bibnamefont {Lang}}, \bibinfo {author} {\bibfnamefont {J.}~\bibnamefont
			{Neuefeind}}, \bibinfo {author} {\bibfnamefont {J.~A.}\ \bibnamefont
			{Alonso}},\ and\ \bibinfo {author} {\bibfnamefont {S.}~\bibnamefont
			{Johnston}},\ }\bibfield  {title} {\bibinfo {title} {Experimental evidence
			for bipolaron condensation as a mechanism for the metal-insulator transition
			in rare-earth nickelates},\ }\href
	{https://doi.org/10.1038/s41467-017-02561-6} {\bibfield  {journal} {\bibinfo
			{journal} {Nature Communications}\ }\textbf {\bibinfo {volume} {9}},\
		\bibinfo {pages} {86} (\bibinfo {year} {2018})}\BibitemShut {NoStop}%
	\bibitem [{\citenamefont {Shi}\ \emph {et~al.}(2014)\citenamefont {Shi},
		\citenamefont {Zhou},\ and\ \citenamefont {Ramanathan}}]{Shi:2014p4860}%
	\BibitemOpen
	\bibfield  {author} {\bibinfo {author} {\bibfnamefont {J.}~\bibnamefont
			{Shi}}, \bibinfo {author} {\bibfnamefont {Y.}~\bibnamefont {Zhou}},\ and\
		\bibinfo {author} {\bibfnamefont {S.}~\bibnamefont {Ramanathan}},\ }\bibfield
	{title} {\bibinfo {title} {Colossal resistance switching and band gap
			modulation in a perovskite nickelate by electron doping},\ }\href
	{https://doi.org/10.1038/ncomms5860} {\bibfield  {journal} {\bibinfo
			{journal} {Nature Communications}\ }\textbf {\bibinfo {volume} {5}},\
		\bibinfo {pages} {4860} (\bibinfo {year} {2014})}\BibitemShut {NoStop}%
	\bibitem [{\citenamefont {Kotiuga}\ \emph {et~al.}(2019)\citenamefont
		{Kotiuga}, \citenamefont {Zhang}, \citenamefont {Li}, \citenamefont
		{Rodolakis}, \citenamefont {Zhou}, \citenamefont {Sutarto}, \citenamefont
		{He}, \citenamefont {Wang}, \citenamefont {Sun}, \citenamefont {Wang},
		\citenamefont {Aghamiri}, \citenamefont {Hancock}, \citenamefont {Rokhinson},
		\citenamefont {Landau}, \citenamefont {Abate}, \citenamefont {Freeland},
		\citenamefont {Comin}, \citenamefont {Ramanathan},\ and\ \citenamefont
		{Rabe}}]{Kotiuga:2019p21992}%
	\BibitemOpen
	\bibfield  {author} {\bibinfo {author} {\bibfnamefont {M.}~\bibnamefont
			{Kotiuga}}, \bibinfo {author} {\bibfnamefont {Z.}~\bibnamefont {Zhang}},
		\bibinfo {author} {\bibfnamefont {J.}~\bibnamefont {Li}}, \bibinfo {author}
		{\bibfnamefont {F.}~\bibnamefont {Rodolakis}}, \bibinfo {author}
		{\bibfnamefont {H.}~\bibnamefont {Zhou}}, \bibinfo {author} {\bibfnamefont
			{R.}~\bibnamefont {Sutarto}}, \bibinfo {author} {\bibfnamefont
			{F.}~\bibnamefont {He}}, \bibinfo {author} {\bibfnamefont {Q.}~\bibnamefont
			{Wang}}, \bibinfo {author} {\bibfnamefont {Y.}~\bibnamefont {Sun}}, \bibinfo
		{author} {\bibfnamefont {Y.}~\bibnamefont {Wang}}, \bibinfo {author}
		{\bibfnamefont {N.~A.}\ \bibnamefont {Aghamiri}}, \bibinfo {author}
		{\bibfnamefont {S.~B.}\ \bibnamefont {Hancock}}, \bibinfo {author}
		{\bibfnamefont {L.~P.}\ \bibnamefont {Rokhinson}}, \bibinfo {author}
		{\bibfnamefont {D.~P.}\ \bibnamefont {Landau}}, \bibinfo {author}
		{\bibfnamefont {Y.}~\bibnamefont {Abate}}, \bibinfo {author} {\bibfnamefont
			{J.~W.}\ \bibnamefont {Freeland}}, \bibinfo {author} {\bibfnamefont
			{R.}~\bibnamefont {Comin}}, \bibinfo {author} {\bibfnamefont
			{S.}~\bibnamefont {Ramanathan}},\ and\ \bibinfo {author} {\bibfnamefont
			{K.~M.}\ \bibnamefont {Rabe}},\ }\bibfield  {title} {\bibinfo {title}
		{Carrier localization in perovskite nickelates from oxygen vacancies},\
	}\href {https://doi.org/10.1073/pnas.1910490116} {\bibfield  {journal}
		{\bibinfo  {journal} {Proceedings of the National Academy of Sciences}\
		}\textbf {\bibinfo {volume} {116}},\ \bibinfo {pages} {21992} (\bibinfo
		{year} {2019})}\BibitemShut {NoStop}%
	\bibitem [{\citenamefont {Li}\ \emph {et~al.}(2021)\citenamefont {Li},
		\citenamefont {Green}, \citenamefont {Zhang}, \citenamefont {Sutarto},
		\citenamefont {Sadowski}, \citenamefont {Zhu}, \citenamefont {Zhang},
		\citenamefont {Zhou}, \citenamefont {Sun}, \citenamefont {He}, \citenamefont
		{Ramanathan},\ and\ \citenamefont {Comin}}]{Li:2021p187602}%
	\BibitemOpen
	\bibfield  {author} {\bibinfo {author} {\bibfnamefont {J.}~\bibnamefont
			{Li}}, \bibinfo {author} {\bibfnamefont {R.~J.}\ \bibnamefont {Green}},
		\bibinfo {author} {\bibfnamefont {Z.}~\bibnamefont {Zhang}}, \bibinfo
		{author} {\bibfnamefont {R.}~\bibnamefont {Sutarto}}, \bibinfo {author}
		{\bibfnamefont {J.~T.}\ \bibnamefont {Sadowski}}, \bibinfo {author}
		{\bibfnamefont {Z.}~\bibnamefont {Zhu}}, \bibinfo {author} {\bibfnamefont
			{G.}~\bibnamefont {Zhang}}, \bibinfo {author} {\bibfnamefont
			{D.}~\bibnamefont {Zhou}}, \bibinfo {author} {\bibfnamefont {Y.}~\bibnamefont
			{Sun}}, \bibinfo {author} {\bibfnamefont {F.}~\bibnamefont {He}}, \bibinfo
		{author} {\bibfnamefont {S.}~\bibnamefont {Ramanathan}},\ and\ \bibinfo
		{author} {\bibfnamefont {R.}~\bibnamefont {Comin}},\ }\bibfield  {title}
	{\bibinfo {title} {Sudden collapse of magnetic order in oxygen-deficient
			nickelate films},\ }\href {https://doi.org/10.1103/PhysRevLett.126.187602}
	{\bibfield  {journal} {\bibinfo  {journal} {Phys. Rev. Lett.}\ }\textbf
		{\bibinfo {volume} {126}},\ \bibinfo {pages} {187602} (\bibinfo {year}
		{2021})}\BibitemShut {NoStop}%
	\bibitem [{\citenamefont {Liu}\ \emph {et~al.}(2019)\citenamefont {Liu},
		\citenamefont {Dalpian},\ and\ \citenamefont {Zunger}}]{Liu:2019p106403}%
	\BibitemOpen
	\bibfield  {author} {\bibinfo {author} {\bibfnamefont {Q.}~\bibnamefont
			{Liu}}, \bibinfo {author} {\bibfnamefont {G.~M.}\ \bibnamefont {Dalpian}},\
		and\ \bibinfo {author} {\bibfnamefont {A.}~\bibnamefont {Zunger}},\
	}\bibfield  {title} {\bibinfo {title} {Antidoping in insulators and
			semiconductors having intermediate bands with trapped carriers},\ }\href
	{https://doi.org/10.1103/PhysRevLett.122.106403} {\bibfield  {journal}
		{\bibinfo  {journal} {Phys. Rev. Lett.}\ }\textbf {\bibinfo {volume} {122}},\
		\bibinfo {pages} {106403} (\bibinfo {year} {2019})}\BibitemShut {NoStop}%
	\bibitem [{\citenamefont {Shi}\ \emph {et~al.}(2024)\citenamefont {Shi},
		\citenamefont {Zhao}, \citenamefont {Dabo}, \citenamefont {Ramanathan},\ and\
		\citenamefont {Chen}}]{Shi:2024p256502}%
	\BibitemOpen
	\bibfield  {author} {\bibinfo {author} {\bibfnamefont {Y.}~\bibnamefont
			{Shi}}, \bibinfo {author} {\bibfnamefont {G.-D.}\ \bibnamefont {Zhao}},
		\bibinfo {author} {\bibfnamefont {I.}~\bibnamefont {Dabo}}, \bibinfo {author}
		{\bibfnamefont {S.}~\bibnamefont {Ramanathan}},\ and\ \bibinfo {author}
		{\bibfnamefont {L.-Q.}\ \bibnamefont {Chen}},\ }\bibfield  {title} {\bibinfo
		{title} {Phase-field model of electronic antidoping},\ }\href
	{https://doi.org/10.1103/PhysRevLett.132.256502} {\bibfield  {journal}
		{\bibinfo  {journal} {Phys. Rev. Lett.}\ }\textbf {\bibinfo {volume} {132}},\
		\bibinfo {pages} {256502} (\bibinfo {year} {2024})}\BibitemShut {NoStop}%
	\bibitem [{\citenamefont {Aamlid}\ \emph {et~al.}(2023)\citenamefont {Aamlid},
		\citenamefont {Oudah}, \citenamefont {Rottler},\ and\ \citenamefont
		{Hallas}}]{Aamlid:2023p5991}%
	\BibitemOpen
	\bibfield  {author} {\bibinfo {author} {\bibfnamefont {S.~S.}\ \bibnamefont
			{Aamlid}}, \bibinfo {author} {\bibfnamefont {M.}~\bibnamefont {Oudah}},
		\bibinfo {author} {\bibfnamefont {J.}~\bibnamefont {Rottler}},\ and\ \bibinfo
		{author} {\bibfnamefont {A.~M.}\ \bibnamefont {Hallas}},\ }\bibfield  {title}
	{\bibinfo {title} {Understanding the role of entropy in high entropy
			oxides},\ }\href@noop {} {\bibfield  {journal} {\bibinfo  {journal} {Journal
				of the American Chemical Society}\ }\textbf {\bibinfo {volume} {145}},\
		\bibinfo {pages} {5991} (\bibinfo {year} {2023})}\BibitemShut {NoStop}%
	\bibitem [{\citenamefont {Guo}\ and\ \citenamefont
		{Noheda}(2021)}]{Guo:2021p72}%
	\BibitemOpen
	\bibfield  {author} {\bibinfo {author} {\bibfnamefont {Q.}~\bibnamefont
			{Guo}}\ and\ \bibinfo {author} {\bibfnamefont {B.}~\bibnamefont {Noheda}},\
	}\bibfield  {title} {\bibinfo {title} {From hidden metal-insulator transition
			to planckian-like dissipation by tuning the oxygen content in a nickelate},\
	}\href {https://doi.org/10.1038/s41535-021-00374-x} {\bibfield  {journal}
		{\bibinfo  {journal} {npj Quantum Materials}\ }\textbf {\bibinfo {volume}
			{6}},\ \bibinfo {pages} {72} (\bibinfo {year} {2021})}\BibitemShut {NoStop}%
	\bibitem [{\citenamefont {Jeong}\ \emph {et~al.}(2013)\citenamefont {Jeong},
		\citenamefont {Aetukuri}, \citenamefont {Graf}, \citenamefont {Schladt},
		\citenamefont {Samant},\ and\ \citenamefont {Parkin}}]{Jeong:2013p14021405}%
	\BibitemOpen
	\bibfield  {author} {\bibinfo {author} {\bibfnamefont {J.}~\bibnamefont
			{Jeong}}, \bibinfo {author} {\bibfnamefont {N.}~\bibnamefont {Aetukuri}},
		\bibinfo {author} {\bibfnamefont {T.}~\bibnamefont {Graf}}, \bibinfo {author}
		{\bibfnamefont {T.~D.}\ \bibnamefont {Schladt}}, \bibinfo {author}
		{\bibfnamefont {M.~G.}\ \bibnamefont {Samant}},\ and\ \bibinfo {author}
		{\bibfnamefont {S.~S.~P.}\ \bibnamefont {Parkin}},\ }\bibfield  {title}
	{\bibinfo {title} {Suppression of metal-insulator transition in
			vo<sub>2</sub> by electric field\&\#x2013;induced oxygen vacancy formation},\
	}\href {https://doi.org/10.1126/science.1230512} {\bibfield  {journal}
		{\bibinfo  {journal} {Science}\ }\textbf {\bibinfo {volume} {339}},\ \bibinfo
		{pages} {1402} (\bibinfo {year} {2013})}\BibitemShut {NoStop}%
	\bibitem [{\citenamefont {Park}\ \emph {et~al.}(2020)\citenamefont {Park},
		\citenamefont {Sim}, \citenamefont {Jo}, \citenamefont {Kim}, \citenamefont
		{Yoon}, \citenamefont {Han}, \citenamefont {Kim}, \citenamefont {Song},
		\citenamefont {Lee}, \citenamefont {Choi},\ and\ \citenamefont
		{Son}}]{Park:2020P1401}%
	\BibitemOpen
	\bibfield  {author} {\bibinfo {author} {\bibfnamefont {Y.}~\bibnamefont
			{Park}}, \bibinfo {author} {\bibfnamefont {H.}~\bibnamefont {Sim}}, \bibinfo
		{author} {\bibfnamefont {M.}~\bibnamefont {Jo}}, \bibinfo {author}
		{\bibfnamefont {G.-Y.}\ \bibnamefont {Kim}}, \bibinfo {author} {\bibfnamefont
			{D.}~\bibnamefont {Yoon}}, \bibinfo {author} {\bibfnamefont {H.}~\bibnamefont
			{Han}}, \bibinfo {author} {\bibfnamefont {Y.}~\bibnamefont {Kim}}, \bibinfo
		{author} {\bibfnamefont {K.}~\bibnamefont {Song}}, \bibinfo {author}
		{\bibfnamefont {D.}~\bibnamefont {Lee}}, \bibinfo {author} {\bibfnamefont
			{S.-Y.}\ \bibnamefont {Choi}},\ and\ \bibinfo {author} {\bibfnamefont
			{J.}~\bibnamefont {Son}},\ }\bibfield  {title} {\bibinfo {title} {Directional
			ionic transport across the oxide interface enables low-temperature epitaxy of
			rutile tio2},\ }\href {https://doi.org/10.1038/s41467-020-15142-x} {\bibfield
		{journal} {\bibinfo  {journal} {Nature Communications}\ }\textbf {\bibinfo
			{volume} {11}},\ \bibinfo {pages} {1401} (\bibinfo {year}
		{2020})}\BibitemShut {NoStop}%
	\bibitem [{\citenamefont {Brockman}\ \emph {et~al.}(2011)\citenamefont
		{Brockman}, \citenamefont {Aetukuri}, \citenamefont {Topuria}, \citenamefont
		{Samant}, \citenamefont {Roche},\ and\ \citenamefont
		{Parkin}}]{Brockman:2011p152105}%
	\BibitemOpen
	\bibfield  {author} {\bibinfo {author} {\bibfnamefont {J.}~\bibnamefont
			{Brockman}}, \bibinfo {author} {\bibfnamefont {N.~P.}\ \bibnamefont
			{Aetukuri}}, \bibinfo {author} {\bibfnamefont {T.}~\bibnamefont {Topuria}},
		\bibinfo {author} {\bibfnamefont {M.~G.}\ \bibnamefont {Samant}}, \bibinfo
		{author} {\bibfnamefont {K.~P.}\ \bibnamefont {Roche}},\ and\ \bibinfo
		{author} {\bibfnamefont {S.~S.~P.}\ \bibnamefont {Parkin}},\ }\bibfield
	{title} {\bibinfo {title} {Increased metal-insulator transition temperatures
			in epitaxial thin films of v2o3 prepared in reduced oxygen environments},\
	}\href {https://doi.org/10.1063/1.3574910} {\bibfield  {journal} {\bibinfo
			{journal} {Applied Physics Letters}\ }\textbf {\bibinfo {volume} {98}},\
		\bibinfo {pages} {152105} (\bibinfo {year} {2011})}\BibitemShut {NoStop}%
	\bibitem [{\citenamefont {Gorbenko}\ \emph {et~al.}(2002)\citenamefont
		{Gorbenko}, \citenamefont {Samoilenkov}, \citenamefont {Graboy},\ and\
		\citenamefont {Kaul}}]{Gorbenko:2002p4026}%
	\BibitemOpen
	\bibfield  {author} {\bibinfo {author} {\bibfnamefont {O.~Y.}\ \bibnamefont
			{Gorbenko}}, \bibinfo {author} {\bibfnamefont {S.}~\bibnamefont
			{Samoilenkov}}, \bibinfo {author} {\bibfnamefont {I.}~\bibnamefont
			{Graboy}},\ and\ \bibinfo {author} {\bibfnamefont {A.}~\bibnamefont {Kaul}},\
	}\bibfield  {title} {\bibinfo {title} {Epitaxial stabilization of oxides in
			thin films},\ }\href@noop {} {\bibfield  {journal} {\bibinfo  {journal}
			{Chemistry of materials}\ }\textbf {\bibinfo {volume} {14}},\ \bibinfo
		{pages} {4026} (\bibinfo {year} {2002})}\BibitemShut {NoStop}%
	\bibitem [{\citenamefont {Wei}\ \emph {et~al.}(2023)\citenamefont {Wei},
		\citenamefont {Vu}, \citenamefont {Zhang}, \citenamefont {Walker},\ and\
		\citenamefont {Ahn}}]{Wei:2023peadh3327}%
	\BibitemOpen
	\bibfield  {author} {\bibinfo {author} {\bibfnamefont {W.}~\bibnamefont
			{Wei}}, \bibinfo {author} {\bibfnamefont {D.}~\bibnamefont {Vu}}, \bibinfo
		{author} {\bibfnamefont {Z.}~\bibnamefont {Zhang}}, \bibinfo {author}
		{\bibfnamefont {F.~J.}\ \bibnamefont {Walker}},\ and\ \bibinfo {author}
		{\bibfnamefont {C.~H.}\ \bibnamefont {Ahn}},\ }\bibfield  {title} {\bibinfo
		{title} {Superconducting nd1- x eu x nio2 thin films using in situ
			synthesis},\ }\href@noop {} {\bibfield  {journal} {\bibinfo  {journal}
			{Science Advances}\ }\textbf {\bibinfo {volume} {9}},\ \bibinfo {pages}
		{eadh3327} (\bibinfo {year} {2023})}\BibitemShut {NoStop}%
	\bibitem [{\citenamefont {Bhattacharya}\ \emph {et~al.}()\citenamefont
		{Bhattacharya}, \citenamefont {Joshi}, \citenamefont {Patel}, \citenamefont
		{Zhang}, \citenamefont {Saha}, \citenamefont {Mandal}, \citenamefont {Ojha},
		\citenamefont {Gloskovskii}, \citenamefont {Schlueter}, \citenamefont
		{Freeland}, \citenamefont {Zhang}, \citenamefont {Zhou}, \citenamefont
		{Yang},\ and\ \citenamefont {Middey}}]{Bhattacharya:2025p2418490}%
	\BibitemOpen
	\bibfield  {author} {\bibinfo {author} {\bibfnamefont {N.}~\bibnamefont
			{Bhattacharya}}, \bibinfo {author} {\bibfnamefont {S.~C.}\ \bibnamefont
			{Joshi}}, \bibinfo {author} {\bibfnamefont {R.~K.}\ \bibnamefont {Patel}},
		\bibinfo {author} {\bibfnamefont {J.}~\bibnamefont {Zhang}}, \bibinfo
		{author} {\bibfnamefont {A.}~\bibnamefont {Saha}}, \bibinfo {author}
		{\bibfnamefont {P.}~\bibnamefont {Mandal}}, \bibinfo {author} {\bibfnamefont
			{S.~K.}\ \bibnamefont {Ojha}}, \bibinfo {author} {\bibfnamefont
			{A.}~\bibnamefont {Gloskovskii}}, \bibinfo {author} {\bibfnamefont
			{C.}~\bibnamefont {Schlueter}}, \bibinfo {author} {\bibfnamefont {J.~W.}\
			\bibnamefont {Freeland}}, \bibinfo {author} {\bibfnamefont {Z.}~\bibnamefont
			{Zhang}}, \bibinfo {author} {\bibfnamefont {H.}~\bibnamefont {Zhou}},
		\bibinfo {author} {\bibfnamefont {Z.}~\bibnamefont {Yang}},\ and\ \bibinfo
		{author} {\bibfnamefont {S.}~\bibnamefont {Middey}},\ }\bibfield  {title}
	{\bibinfo {title} {Nanoscale inhomogeneity and epitaxial strain control
			metallicity in single crystalline thin films of high entropy oxide},\ }\href
	{https://doi.org/https://doi.org/10.1002/adma.202418490} {\bibfield
		{journal} {\bibinfo  {journal} {Advanced Materials}\ }\textbf {\bibinfo
			{volume} {n/a}},\ \bibinfo {pages} {2418490}}\BibitemShut {NoStop}%
	\bibitem [{\citenamefont {Abbate}\ \emph {et~al.}(2002)\citenamefont {Abbate},
		\citenamefont {Zampieri}, \citenamefont {Prado}, \citenamefont {Caneiro},
		\citenamefont {Gonzalez-Calbet},\ and\ \citenamefont
		{Vallet-Regi}}]{Abbate:2002p155101}%
	\BibitemOpen
	\bibfield  {author} {\bibinfo {author} {\bibfnamefont {M.}~\bibnamefont
			{Abbate}}, \bibinfo {author} {\bibfnamefont {G.}~\bibnamefont {Zampieri}},
		\bibinfo {author} {\bibfnamefont {F.}~\bibnamefont {Prado}}, \bibinfo
		{author} {\bibfnamefont {A.}~\bibnamefont {Caneiro}}, \bibinfo {author}
		{\bibfnamefont {J.~M.}\ \bibnamefont {Gonzalez-Calbet}},\ and\ \bibinfo
		{author} {\bibfnamefont {M.}~\bibnamefont {Vallet-Regi}},\ }\bibfield
	{title} {\bibinfo {title} {Electronic structure and metal-insulator
			transition in ${\mathrm{lanio}}_{3\ensuremath{-}\ensuremath{\delta}}$},\
	}\href {https://doi.org/10.1103/PhysRevB.65.155101} {\bibfield  {journal}
		{\bibinfo  {journal} {Phys. Rev. B}\ }\textbf {\bibinfo {volume} {65}},\
		\bibinfo {pages} {155101} (\bibinfo {year} {2002})}\BibitemShut {NoStop}%
	\bibitem [{\citenamefont {Middey}\ \emph {et~al.}(2014)\citenamefont {Middey},
		\citenamefont {Rivero}, \citenamefont {Meyers}, \citenamefont {Kareev},
		\citenamefont {Liu}, \citenamefont {Cao}, \citenamefont {Freeland},
		\citenamefont {Barraza-Lopez},\ and\ \citenamefont
		{Chakhalian}}]{Middey:2014p6819}%
	\BibitemOpen
	\bibfield  {author} {\bibinfo {author} {\bibfnamefont {S.}~\bibnamefont
			{Middey}}, \bibinfo {author} {\bibfnamefont {P.}~\bibnamefont {Rivero}},
		\bibinfo {author} {\bibfnamefont {D.}~\bibnamefont {Meyers}}, \bibinfo
		{author} {\bibfnamefont {M.}~\bibnamefont {Kareev}}, \bibinfo {author}
		{\bibfnamefont {X.}~\bibnamefont {Liu}}, \bibinfo {author} {\bibfnamefont
			{Y.}~\bibnamefont {Cao}}, \bibinfo {author} {\bibfnamefont {J.}~\bibnamefont
			{Freeland}}, \bibinfo {author} {\bibfnamefont {S.}~\bibnamefont
			{Barraza-Lopez}},\ and\ \bibinfo {author} {\bibfnamefont {J.}~\bibnamefont
			{Chakhalian}},\ }\bibfield  {title} {\bibinfo {title} {Polarity compensation
			in ultra-thin films of complex oxides: The case of a perovskite nickelate},\
	}\href@noop {} {\bibfield  {journal} {\bibinfo  {journal} {Scientific
				reports}\ }\textbf {\bibinfo {volume} {4}},\ \bibinfo {pages} {6819}
		(\bibinfo {year} {2014})}\BibitemShut {NoStop}%
	\bibitem [{\citenamefont {Chakhalian}\ \emph {et~al.}(2011)\citenamefont
		{Chakhalian}, \citenamefont {Rondinelli}, \citenamefont {Liu}, \citenamefont
		{Gray}, \citenamefont {Kareev}, \citenamefont {Moon}, \citenamefont {Prasai},
		\citenamefont {Cohn}, \citenamefont {Varela}, \citenamefont {Tung},
		\citenamefont {Bedzyk}, \citenamefont {Altendorf}, \citenamefont {Strigari},
		\citenamefont {Dabrowski}, \citenamefont {Tjeng}, \citenamefont {Ryan},\ and\
		\citenamefont {Freeland}}]{Chakhalian:2011p116805}%
	\BibitemOpen
	\bibfield  {author} {\bibinfo {author} {\bibfnamefont {J.}~\bibnamefont
			{Chakhalian}}, \bibinfo {author} {\bibfnamefont {J.~M.}\ \bibnamefont
			{Rondinelli}}, \bibinfo {author} {\bibfnamefont {J.}~\bibnamefont {Liu}},
		\bibinfo {author} {\bibfnamefont {B.~A.}\ \bibnamefont {Gray}}, \bibinfo
		{author} {\bibfnamefont {M.}~\bibnamefont {Kareev}}, \bibinfo {author}
		{\bibfnamefont {E.~J.}\ \bibnamefont {Moon}}, \bibinfo {author}
		{\bibfnamefont {N.}~\bibnamefont {Prasai}}, \bibinfo {author} {\bibfnamefont
			{J.~L.}\ \bibnamefont {Cohn}}, \bibinfo {author} {\bibfnamefont
			{M.}~\bibnamefont {Varela}}, \bibinfo {author} {\bibfnamefont {I.~C.}\
			\bibnamefont {Tung}}, \bibinfo {author} {\bibfnamefont {M.~J.}\ \bibnamefont
			{Bedzyk}}, \bibinfo {author} {\bibfnamefont {S.~G.}\ \bibnamefont
			{Altendorf}}, \bibinfo {author} {\bibfnamefont {F.}~\bibnamefont {Strigari}},
		\bibinfo {author} {\bibfnamefont {B.}~\bibnamefont {Dabrowski}}, \bibinfo
		{author} {\bibfnamefont {L.~H.}\ \bibnamefont {Tjeng}}, \bibinfo {author}
		{\bibfnamefont {P.~J.}\ \bibnamefont {Ryan}},\ and\ \bibinfo {author}
		{\bibfnamefont {J.~W.}\ \bibnamefont {Freeland}},\ }\bibfield  {title}
	{\bibinfo {title} {Asymmetric orbital-lattice interactions in ultrathin
			correlated oxide films},\ }\href
	{https://doi.org/10.1103/PhysRevLett.107.116805} {\bibfield  {journal}
		{\bibinfo  {journal} {Phys. Rev. Lett.}\ }\textbf {\bibinfo {volume} {107}},\
		\bibinfo {pages} {116805} (\bibinfo {year} {2011})}\BibitemShut {NoStop}%
	\bibitem [{\citenamefont {Liu}\ \emph {et~al.}(2013)\citenamefont {Liu},
		\citenamefont {Kargarian}, \citenamefont {Kareev}, \citenamefont {Gray},
		\citenamefont {Ryan}, \citenamefont {Cruz}, \citenamefont {Tahir},
		\citenamefont {Chuang}, \citenamefont {Guo}, \citenamefont {Rondinelli},
		\citenamefont {Freeland}, \citenamefont {Fiete},\ and\ \citenamefont
		{Chakhalian}}]{Liu:2013p2714}%
	\BibitemOpen
	\bibfield  {author} {\bibinfo {author} {\bibfnamefont {J.}~\bibnamefont
			{Liu}}, \bibinfo {author} {\bibfnamefont {M.}~\bibnamefont {Kargarian}},
		\bibinfo {author} {\bibfnamefont {M.}~\bibnamefont {Kareev}}, \bibinfo
		{author} {\bibfnamefont {B.}~\bibnamefont {Gray}}, \bibinfo {author}
		{\bibfnamefont {P.~J.}\ \bibnamefont {Ryan}}, \bibinfo {author}
		{\bibfnamefont {A.}~\bibnamefont {Cruz}}, \bibinfo {author} {\bibfnamefont
			{N.}~\bibnamefont {Tahir}}, \bibinfo {author} {\bibfnamefont {Y.-D.}\
			\bibnamefont {Chuang}}, \bibinfo {author} {\bibfnamefont {J.}~\bibnamefont
			{Guo}}, \bibinfo {author} {\bibfnamefont {J.~M.}\ \bibnamefont {Rondinelli}},
		\bibinfo {author} {\bibfnamefont {J.~W.}\ \bibnamefont {Freeland}}, \bibinfo
		{author} {\bibfnamefont {G.~A.}\ \bibnamefont {Fiete}},\ and\ \bibinfo
		{author} {\bibfnamefont {J.}~\bibnamefont {Chakhalian}},\ }\bibfield  {title}
	{\bibinfo {title} {Heterointerface engineered electronic and magnetic phases
			of ndnio3 thin films},\ }\href {https://doi.org/10.1038/ncomms3714}
	{\bibfield  {journal} {\bibinfo  {journal} {Nature Communications}\ }\textbf
		{\bibinfo {volume} {4}},\ \bibinfo {pages} {2714} (\bibinfo {year}
		{2013})}\BibitemShut {NoStop}%
	\bibitem [{\citenamefont {Middey}\ \emph
		{et~al.}(2018{\natexlab{b}})\citenamefont {Middey}, \citenamefont {Meyers},
		\citenamefont {Ojha}, \citenamefont {Kareev}, \citenamefont {Liu},
		\citenamefont {Cao}, \citenamefont {Freeland},\ and\ \citenamefont
		{Chakhalian}}]{Middey:2018p045115}%
	\BibitemOpen
	\bibfield  {author} {\bibinfo {author} {\bibfnamefont {S.}~\bibnamefont
			{Middey}}, \bibinfo {author} {\bibfnamefont {D.}~\bibnamefont {Meyers}},
		\bibinfo {author} {\bibfnamefont {S.~K.}\ \bibnamefont {Ojha}}, \bibinfo
		{author} {\bibfnamefont {M.}~\bibnamefont {Kareev}}, \bibinfo {author}
		{\bibfnamefont {X.}~\bibnamefont {Liu}}, \bibinfo {author} {\bibfnamefont
			{Y.}~\bibnamefont {Cao}}, \bibinfo {author} {\bibfnamefont {J.~W.}\
			\bibnamefont {Freeland}},\ and\ \bibinfo {author} {\bibfnamefont
			{J.}~\bibnamefont {Chakhalian}},\ }\bibfield  {title} {\bibinfo {title}
		{Epitaxial strain modulated electronic properties of interface controlled
			nickelate superlattices},\ }\href
	{https://doi.org/10.1103/PhysRevB.98.045115} {\bibfield  {journal} {\bibinfo
			{journal} {Phys. Rev. B}\ }\textbf {\bibinfo {volume} {98}},\ \bibinfo
		{pages} {045115} (\bibinfo {year} {2018}{\natexlab{b}})}\BibitemShut
	{NoStop}%
	\bibitem [{\citenamefont {Horiba}\ \emph {et~al.}(2007)\citenamefont {Horiba},
		\citenamefont {Eguchi}, \citenamefont {Taguchi}, \citenamefont {Chainani},
		\citenamefont {Kikkawa}, \citenamefont {Senba}, \citenamefont {Ohashi},\ and\
		\citenamefont {Shin}}]{Horiba:2007p155104}%
	\BibitemOpen
	\bibfield  {author} {\bibinfo {author} {\bibfnamefont {K.}~\bibnamefont
			{Horiba}}, \bibinfo {author} {\bibfnamefont {R.}~\bibnamefont {Eguchi}},
		\bibinfo {author} {\bibfnamefont {M.}~\bibnamefont {Taguchi}}, \bibinfo
		{author} {\bibfnamefont {A.}~\bibnamefont {Chainani}}, \bibinfo {author}
		{\bibfnamefont {A.}~\bibnamefont {Kikkawa}}, \bibinfo {author} {\bibfnamefont
			{Y.}~\bibnamefont {Senba}}, \bibinfo {author} {\bibfnamefont
			{H.}~\bibnamefont {Ohashi}},\ and\ \bibinfo {author} {\bibfnamefont
			{S.}~\bibnamefont {Shin}},\ }\bibfield  {title} {\bibinfo {title} {Electronic
			structure of $\mathrm{La}\mathrm{Ni}{\mathrm{o}}_{3\ensuremath{-}x}$: An in
			situ soft x-ray photoemission and absorption study},\ }\href
	{https://doi.org/10.1103/PhysRevB.76.155104} {\bibfield  {journal} {\bibinfo
			{journal} {Phys. Rev. B}\ }\textbf {\bibinfo {volume} {76}},\ \bibinfo
		{pages} {155104} (\bibinfo {year} {2007})}\BibitemShut {NoStop}%
	\bibitem [{\citenamefont {Guo}\ \emph {et~al.}(2020)\citenamefont {Guo},
		\citenamefont {Farokhipoor}, \citenamefont {Mag{\'e}n}, \citenamefont
		{Rivadulla},\ and\ \citenamefont {Noheda}}]{Guo:2020p2949}%
	\BibitemOpen
	\bibfield  {author} {\bibinfo {author} {\bibfnamefont {Q.}~\bibnamefont
			{Guo}}, \bibinfo {author} {\bibfnamefont {S.}~\bibnamefont {Farokhipoor}},
		\bibinfo {author} {\bibfnamefont {C.}~\bibnamefont {Mag{\'e}n}}, \bibinfo
		{author} {\bibfnamefont {F.}~\bibnamefont {Rivadulla}},\ and\ \bibinfo
		{author} {\bibfnamefont {B.}~\bibnamefont {Noheda}},\ }\bibfield  {title}
	{\bibinfo {title} {Tunable resistivity exponents in the metallic phase of
			epitaxial nickelates},\ }\href@noop {} {\bibfield  {journal} {\bibinfo
			{journal} {Nature Communications}\ }\textbf {\bibinfo {volume} {11}},\
		\bibinfo {pages} {2949} (\bibinfo {year} {2020})}\BibitemShut {NoStop}%
	\bibitem [{\citenamefont {S\'anchez}\ \emph {et~al.}(1996)\citenamefont
		{S\'anchez}, \citenamefont {Causa}, \citenamefont {Caneiro}, \citenamefont
		{Butera}, \citenamefont {Vallet-Reg\'{\i}}, \citenamefont {Sayagu\'es},
		\citenamefont {Gonz\'alez-Calbet}, \citenamefont {Garc\'{\i}a-Sanz},\ and\
		\citenamefont {Rivas}}]{Sanchez:1996p16574}%
	\BibitemOpen
	\bibfield  {author} {\bibinfo {author} {\bibfnamefont {R.~D.}\ \bibnamefont
			{S\'anchez}}, \bibinfo {author} {\bibfnamefont {M.~T.}\ \bibnamefont
			{Causa}}, \bibinfo {author} {\bibfnamefont {A.}~\bibnamefont {Caneiro}},
		\bibinfo {author} {\bibfnamefont {A.}~\bibnamefont {Butera}}, \bibinfo
		{author} {\bibfnamefont {M.}~\bibnamefont {Vallet-Reg\'{\i}}}, \bibinfo
		{author} {\bibfnamefont {M.~J.}\ \bibnamefont {Sayagu\'es}}, \bibinfo
		{author} {\bibfnamefont {J.}~\bibnamefont {Gonz\'alez-Calbet}}, \bibinfo
		{author} {\bibfnamefont {F.}~\bibnamefont {Garc\'{\i}a-Sanz}},\ and\ \bibinfo
		{author} {\bibfnamefont {J.}~\bibnamefont {Rivas}},\ }\bibfield  {title}
	{\bibinfo {title} {Metal-insulator transition in oxygen-deficient
			${\mathrm{lanio}}_{3\mathrm{\ensuremath{-}}\mathit{x}}$ perovskites},\ }\href
	{https://doi.org/10.1103/PhysRevB.54.16574} {\bibfield  {journal} {\bibinfo
			{journal} {Phys. Rev. B}\ }\textbf {\bibinfo {volume} {54}},\ \bibinfo
		{pages} {16574} (\bibinfo {year} {1996})}\BibitemShut {NoStop}%
	\bibitem [{\citenamefont {Iglesias}\ \emph {et~al.}(2021)\citenamefont
		{Iglesias}, \citenamefont {Bibes},\ and\ \citenamefont
		{Varignon}}]{Iglesias:2021p035123}%
	\BibitemOpen
	\bibfield  {author} {\bibinfo {author} {\bibfnamefont {L.}~\bibnamefont
			{Iglesias}}, \bibinfo {author} {\bibfnamefont {M.}~\bibnamefont {Bibes}},\
		and\ \bibinfo {author} {\bibfnamefont {J.}~\bibnamefont {Varignon}},\
	}\bibfield  {title} {\bibinfo {title} {First-principles study of electron and
			hole doping effects in perovskite nickelates},\ }\href
	{https://doi.org/10.1103/PhysRevB.104.035123} {\bibfield  {journal} {\bibinfo
			{journal} {Phys. Rev. B}\ }\textbf {\bibinfo {volume} {104}},\ \bibinfo
		{pages} {035123} (\bibinfo {year} {2021})}\BibitemShut {NoStop}%
	\bibitem [{\citenamefont {Garc\'{\i}a-Mu\~noz}\ \emph
		{et~al.}(1995)\citenamefont {Garc\'{\i}a-Mu\~noz}, \citenamefont {Suaaidi},
		\citenamefont {Mart\'{\i}nez-Lope},\ and\ \citenamefont
		{Alonso}}]{Garc:1995p1356313569}%
	\BibitemOpen
	\bibfield  {author} {\bibinfo {author} {\bibfnamefont {J.~L.}\ \bibnamefont
			{Garc\'{\i}a-Mu\~noz}}, \bibinfo {author} {\bibfnamefont {M.}~\bibnamefont
			{Suaaidi}}, \bibinfo {author} {\bibfnamefont {M.~J.}\ \bibnamefont
			{Mart\'{\i}nez-Lope}},\ and\ \bibinfo {author} {\bibfnamefont {J.~A.}\
			\bibnamefont {Alonso}},\ }\bibfield  {title} {\bibinfo {title} {Influence of
			carrier injection on the metal-insulator transition in electron- and
			hole-doped
			${\mathit{r}}_{1\mathrm{\ensuremath{-}}\mathit{x}}$${\mathit{a}}_{\mathit{x}}$${\mathrm{nio}}_{3}$
			perovskites},\ }\href {https://doi.org/10.1103/PhysRevB.52.13563} {\bibfield
		{journal} {\bibinfo  {journal} {Phys. Rev. B}\ }\textbf {\bibinfo {volume}
			{52}},\ \bibinfo {pages} {13563} (\bibinfo {year} {1995})}\BibitemShut
	{NoStop}%
	\bibitem [{\citenamefont {Li}\ \emph {et~al.}(2022)\citenamefont {Li},
		\citenamefont {Ramanathan},\ and\ \citenamefont
		{Comin}}]{Jiarui:2022p2296424X}%
	\BibitemOpen
	\bibfield  {author} {\bibinfo {author} {\bibfnamefont {J.}~\bibnamefont
			{Li}}, \bibinfo {author} {\bibfnamefont {S.}~\bibnamefont {Ramanathan}},\
		and\ \bibinfo {author} {\bibfnamefont {R.}~\bibnamefont {Comin}},\ }\bibfield
	{title} {\bibinfo {title} {Carrier doping physics of rare earth perovskite
			nickelates renio3},\ }\href
	{https://www.frontiersin.org/articles/10.3389/fphy.2022.834882} {\bibfield
		{journal} {\bibinfo  {journal} {Frontiers in Physics}\ }\textbf {\bibinfo
			{volume} {10}} (\bibinfo {year} {2022})}\BibitemShut {NoStop}%
	\bibitem [{\citenamefont {Hadjimichael}\ \emph {et~al.}(2023)\citenamefont
		{Hadjimichael}, \citenamefont {Mundet}, \citenamefont {Domínguez},
		\citenamefont {Waelchli}, \citenamefont {De~Luca}, \citenamefont {Spring},
		\citenamefont {Jöhr}, \citenamefont {McKeown~Walker}, \citenamefont
		{Piamonteze}, \citenamefont {Alexander}, \citenamefont {Triscone},\ and\
		\citenamefont {Gibert}}]{Hadjimichael:2023p2201182}%
	\BibitemOpen
	\bibfield  {author} {\bibinfo {author} {\bibfnamefont {M.}~\bibnamefont
			{Hadjimichael}}, \bibinfo {author} {\bibfnamefont {B.}~\bibnamefont
			{Mundet}}, \bibinfo {author} {\bibfnamefont {C.}~\bibnamefont {Domínguez}},
		\bibinfo {author} {\bibfnamefont {A.}~\bibnamefont {Waelchli}}, \bibinfo
		{author} {\bibfnamefont {G.}~\bibnamefont {De~Luca}}, \bibinfo {author}
		{\bibfnamefont {J.}~\bibnamefont {Spring}}, \bibinfo {author} {\bibfnamefont
			{S.}~\bibnamefont {Jöhr}}, \bibinfo {author} {\bibfnamefont
			{S.}~\bibnamefont {McKeown~Walker}}, \bibinfo {author} {\bibfnamefont
			{C.}~\bibnamefont {Piamonteze}}, \bibinfo {author} {\bibfnamefont {D.~T.~L.}\
			\bibnamefont {Alexander}}, \bibinfo {author} {\bibfnamefont {J.-M.}\
			\bibnamefont {Triscone}},\ and\ \bibinfo {author} {\bibfnamefont
			{M.}~\bibnamefont {Gibert}},\ }\bibfield  {title} {\bibinfo {title}
		{Competition between carrier injection and structural distortions in
			electron-doped perovskite nickelate thin films},\ }\href
	{https://doi.org/https://doi.org/10.1002/aelm.202201182} {\bibfield
		{journal} {\bibinfo  {journal} {Advanced Electronic Materials}\ }\textbf
		{\bibinfo {volume} {9}},\ \bibinfo {pages} {2201182} (\bibinfo {year}
		{2023})}\BibitemShut {NoStop}%
	\bibitem [{\citenamefont {Jaramillo}\ \emph {et~al.}(2014)\citenamefont
		{Jaramillo}, \citenamefont {Ha}, \citenamefont {Silevitch},\ and\
		\citenamefont {Ramanathan}}]{Jaramillo:2014p304307}%
	\BibitemOpen
	\bibfield  {author} {\bibinfo {author} {\bibfnamefont {R.}~\bibnamefont
			{Jaramillo}}, \bibinfo {author} {\bibfnamefont {S.~D.}\ \bibnamefont {Ha}},
		\bibinfo {author} {\bibfnamefont {D.~M.}\ \bibnamefont {Silevitch}},\ and\
		\bibinfo {author} {\bibfnamefont {S.}~\bibnamefont {Ramanathan}},\ }\bibfield
	{title} {\bibinfo {title} {Origins of bad-metal conductivity and the
			insulator--metal transition in the rare-earth nickelates},\ }\href
	{https://doi.org/10.1038/nphys2907} {\bibfield  {journal} {\bibinfo
			{journal} {Nature Physics}\ }\textbf {\bibinfo {volume} {10}},\ \bibinfo
		{pages} {304} (\bibinfo {year} {2014})}\BibitemShut {NoStop}%
	\bibitem [{\citenamefont {Bruin}\ \emph {et~al.}(2013)\citenamefont {Bruin},
		\citenamefont {Sakai}, \citenamefont {Perry},\ and\ \citenamefont
		{Mackenzie}}]{Bruin:2013p804}%
	\BibitemOpen
	\bibfield  {author} {\bibinfo {author} {\bibfnamefont {J.~A.~N.}\
			\bibnamefont {Bruin}}, \bibinfo {author} {\bibfnamefont {H.}~\bibnamefont
			{Sakai}}, \bibinfo {author} {\bibfnamefont {R.~S.}\ \bibnamefont {Perry}},\
		and\ \bibinfo {author} {\bibfnamefont {A.}~\bibnamefont {Mackenzie}},\
	}\bibfield  {title} {\bibinfo {title} {Similarity of scattering rates in
			metals showing t-linear resistivity},\ }\href@noop {} {\bibfield  {journal}
		{\bibinfo  {journal} {Science}\ }\textbf {\bibinfo {volume} {339}},\ \bibinfo
		{pages} {804} (\bibinfo {year} {2013})}\BibitemShut {NoStop}%
	\bibitem [{\citenamefont {Faran}\ and\ \citenamefont
		{Ovadyahu}(1988)}]{Faran:1988p5457}%
	\BibitemOpen
	\bibfield  {author} {\bibinfo {author} {\bibfnamefont {O.}~\bibnamefont
			{Faran}}\ and\ \bibinfo {author} {\bibfnamefont {Z.}~\bibnamefont
			{Ovadyahu}},\ }\bibfield  {title} {\bibinfo {title} {Magnetoconductance in
			the variable-range-hopping regime due to a quantum-interference mechanism},\
	}\href {https://doi.org/10.1103/PhysRevB.38.5457} {\bibfield  {journal}
		{\bibinfo  {journal} {Phys. Rev. B}\ }\textbf {\bibinfo {volume} {38}},\
		\bibinfo {pages} {5457} (\bibinfo {year} {1988})}\BibitemShut {NoStop}%
	\bibitem [{\citenamefont {Mott}(1993)}]{Mott:1993p54575465}%
	\BibitemOpen
	\bibfield  {author} {\bibinfo {author} {\bibfnamefont {S.~N.}\ \bibnamefont
			{Mott}},\ }\href {https://doi.org/10.1093/oso/9780198539797.001.0001} {\emph
		{\bibinfo {title} {{Conduction in Non-Crystalline Materials}}}}\ (\bibinfo
	{publisher} {Oxford University Press},\ \bibinfo {year} {1993})\BibitemShut
	{NoStop}%
	\bibitem [{\citenamefont {Ramadoss}\ \emph {et~al.}(2016)\citenamefont
		{Ramadoss}, \citenamefont {Mandal}, \citenamefont {Dai}, \citenamefont {Wan},
		\citenamefont {Zhou}, \citenamefont {Rokhinson}, \citenamefont {Chen},
		\citenamefont {Hu},\ and\ \citenamefont {Ramanathan}}]{Ramadoss:2016p235124}%
	\BibitemOpen
	\bibfield  {author} {\bibinfo {author} {\bibfnamefont {K.}~\bibnamefont
			{Ramadoss}}, \bibinfo {author} {\bibfnamefont {N.}~\bibnamefont {Mandal}},
		\bibinfo {author} {\bibfnamefont {X.}~\bibnamefont {Dai}}, \bibinfo {author}
		{\bibfnamefont {Z.}~\bibnamefont {Wan}}, \bibinfo {author} {\bibfnamefont
			{Y.}~\bibnamefont {Zhou}}, \bibinfo {author} {\bibfnamefont {L.}~\bibnamefont
			{Rokhinson}}, \bibinfo {author} {\bibfnamefont {Y.~P.}\ \bibnamefont {Chen}},
		\bibinfo {author} {\bibfnamefont {J.}~\bibnamefont {Hu}},\ and\ \bibinfo
		{author} {\bibfnamefont {S.}~\bibnamefont {Ramanathan}},\ }\bibfield  {title}
	{\bibinfo {title} {Sign reversal of magnetoresistance in a perovskite
			nickelate by electron doping},\ }\href
	{https://doi.org/10.1103/PhysRevB.94.235124} {\bibfield  {journal} {\bibinfo
			{journal} {Phys. Rev. B}\ }\textbf {\bibinfo {volume} {94}},\ \bibinfo
		{pages} {235124} (\bibinfo {year} {2016})}\BibitemShut {NoStop}%
	\bibitem [{\citenamefont {Frydman}\ and\ \citenamefont
		{Ovadyahu}(1995)}]{Frydman:1995p745}%
	\BibitemOpen
	\bibfield  {author} {\bibinfo {author} {\bibfnamefont {A.}~\bibnamefont
			{Frydman}}\ and\ \bibinfo {author} {\bibfnamefont {Z.}~\bibnamefont
			{Ovadyahu}},\ }\bibfield  {title} {\bibinfo {title} {Spin and quantum
			interference effects in hopping conductivity},\ }\href
	{https://doi.org/https://doi.org/10.1016/0038-1098(95)00141-7} {\bibfield
		{journal} {\bibinfo  {journal} {Solid State Communications}\ }\textbf
		{\bibinfo {volume} {94}},\ \bibinfo {pages} {745} (\bibinfo {year}
		{1995})}\BibitemShut {NoStop}%
	\bibitem [{\citenamefont {Vaknin}\ \emph {et~al.}(1996)\citenamefont {Vaknin},
		\citenamefont {Frydman}, \citenamefont {Ovadyahu},\ and\ \citenamefont
		{Pollak}}]{Vaknin:1996p13604}%
	\BibitemOpen
	\bibfield  {author} {\bibinfo {author} {\bibfnamefont {A.}~\bibnamefont
			{Vaknin}}, \bibinfo {author} {\bibfnamefont {A.}~\bibnamefont {Frydman}},
		\bibinfo {author} {\bibfnamefont {Z.}~\bibnamefont {Ovadyahu}},\ and\
		\bibinfo {author} {\bibfnamefont {M.}~\bibnamefont {Pollak}},\ }\bibfield
	{title} {\bibinfo {title} {High-field magnetoconductance in anderson
			insulators},\ }\href {https://doi.org/10.1103/PhysRevB.54.13604} {\bibfield
		{journal} {\bibinfo  {journal} {Phys. Rev. B}\ }\textbf {\bibinfo {volume}
			{54}},\ \bibinfo {pages} {13604} (\bibinfo {year} {1996})}\BibitemShut
	{NoStop}%
	\bibitem [{\citenamefont {Sivan}\ \emph {et~al.}(1988)\citenamefont {Sivan},
		\citenamefont {Entin-Wohlman},\ and\ \citenamefont {Imry}}]{Sivan:1988p1566}%
	\BibitemOpen
	\bibfield  {author} {\bibinfo {author} {\bibfnamefont {U.}~\bibnamefont
			{Sivan}}, \bibinfo {author} {\bibfnamefont {O.}~\bibnamefont
			{Entin-Wohlman}},\ and\ \bibinfo {author} {\bibfnamefont {Y.}~\bibnamefont
			{Imry}},\ }\bibfield  {title} {\bibinfo {title} {Orbital magnetoconductance
			in the variable-range--hopping regime},\ }\href
	{https://doi.org/10.1103/PhysRevLett.60.1566} {\bibfield  {journal} {\bibinfo
			{journal} {Phys. Rev. Lett.}\ }\textbf {\bibinfo {volume} {60}},\ \bibinfo
		{pages} {1566} (\bibinfo {year} {1988})}\BibitemShut {NoStop}%
	\bibitem [{\citenamefont {Scherwitzl}\ \emph {et~al.}(2011)\citenamefont
		{Scherwitzl}, \citenamefont {Gariglio}, \citenamefont {Gabay}, \citenamefont
		{Zubko}, \citenamefont {Gibert},\ and\ \citenamefont
		{Triscone}}]{Scherwitzl:2011p246403}%
	\BibitemOpen
	\bibfield  {author} {\bibinfo {author} {\bibfnamefont {R.}~\bibnamefont
			{Scherwitzl}}, \bibinfo {author} {\bibfnamefont {S.}~\bibnamefont
			{Gariglio}}, \bibinfo {author} {\bibfnamefont {M.}~\bibnamefont {Gabay}},
		\bibinfo {author} {\bibfnamefont {P.}~\bibnamefont {Zubko}}, \bibinfo
		{author} {\bibfnamefont {M.}~\bibnamefont {Gibert}},\ and\ \bibinfo {author}
		{\bibfnamefont {J.-M.}\ \bibnamefont {Triscone}},\ }\bibfield  {title}
	{\bibinfo {title} {Metal-insulator transition in ultrathin
			${\mathrm{lanio}}_{3}$ films},\ }\href
	{https://doi.org/10.1103/PhysRevLett.106.246403} {\bibfield  {journal}
		{\bibinfo  {journal} {Phys. Rev. Lett.}\ }\textbf {\bibinfo {volume} {106}},\
		\bibinfo {pages} {246403} (\bibinfo {year} {2011})}\BibitemShut {NoStop}%
	\bibitem [{\citenamefont {Lee}\ and\ \citenamefont
		{Ramakrishnan}(1982)}]{LeeRamakrishnan:1982p40094012}%
	\BibitemOpen
	\bibfield  {author} {\bibinfo {author} {\bibfnamefont {P.~A.}\ \bibnamefont
			{Lee}}\ and\ \bibinfo {author} {\bibfnamefont {T.~V.}\ \bibnamefont
			{Ramakrishnan}},\ }\bibfield  {title} {\bibinfo {title} {Magnetoresistance of
			weakly disordered electrons},\ }\href
	{https://doi.org/10.1103/PhysRevB.26.4009} {\bibfield  {journal} {\bibinfo
			{journal} {Phys. Rev. B}\ }\textbf {\bibinfo {volume} {26}},\ \bibinfo
		{pages} {4009} (\bibinfo {year} {1982})}\BibitemShut {NoStop}%
	\bibitem [{\citenamefont {Bergmann}(1984)}]{BERGMANN:1984p158}%
	\BibitemOpen
	\bibfield  {author} {\bibinfo {author} {\bibfnamefont {G.}~\bibnamefont
			{Bergmann}},\ }\bibfield  {title} {\bibinfo {title} {Weak localization in
			thin films: a time-of-flight experiment with conduction electrons},\ }\href
	{https://doi.org/https://doi.org/10.1016/0370-1573(84)90103-0} {\bibfield
		{journal} {\bibinfo  {journal} {Physics Reports}\ }\textbf {\bibinfo {volume}
			{107}},\ \bibinfo {pages} {1} (\bibinfo {year} {1984})}\BibitemShut {NoStop}%
	\bibitem [{\citenamefont {Liao}\ \emph {et~al.}(2021)\citenamefont {Liao},
		\citenamefont {Singh},\ and\ \citenamefont {Park}}]{Liao:2021p085110}%
	\BibitemOpen
	\bibfield  {author} {\bibinfo {author} {\bibfnamefont {X.}~\bibnamefont
			{Liao}}, \bibinfo {author} {\bibfnamefont {V.}~\bibnamefont {Singh}},\ and\
		\bibinfo {author} {\bibfnamefont {H.}~\bibnamefont {Park}},\ }\bibfield
	{title} {\bibinfo {title} {Oxygen vacancy induced site-selective mott
			transition in ${\mathrm{lanio}}_{3}$},\ }\href
	{https://doi.org/10.1103/PhysRevB.103.085110} {\bibfield  {journal} {\bibinfo
			{journal} {Phys. Rev. B}\ }\textbf {\bibinfo {volume} {103}},\ \bibinfo
		{pages} {085110} (\bibinfo {year} {2021})}\BibitemShut {NoStop}%
	\bibitem [{\citenamefont {Mazza}\ \emph {et~al.}(2023)\citenamefont {Mazza},
		\citenamefont {Acharya}, \citenamefont {Wasik}, \citenamefont {Lapano},
		\citenamefont {Li}, \citenamefont {Musico}, \citenamefont {Keppens},
		\citenamefont {Nelson}, \citenamefont {May}, \citenamefont {Brahlek},
		\citenamefont {Mazzoli}, \citenamefont {Pelliciari}, \citenamefont {Bisogni},
		\citenamefont {Cooper},\ and\ \citenamefont {Ward}}]{Mazza:2023p013008}%
	\BibitemOpen
	\bibfield  {author} {\bibinfo {author} {\bibfnamefont {A.~R.}\ \bibnamefont
			{Mazza}}, \bibinfo {author} {\bibfnamefont {S.~R.}\ \bibnamefont {Acharya}},
		\bibinfo {author} {\bibfnamefont {P.}~\bibnamefont {Wasik}}, \bibinfo
		{author} {\bibfnamefont {J.}~\bibnamefont {Lapano}}, \bibinfo {author}
		{\bibfnamefont {J.}~\bibnamefont {Li}}, \bibinfo {author} {\bibfnamefont
			{B.~L.}\ \bibnamefont {Musico}}, \bibinfo {author} {\bibfnamefont
			{V.}~\bibnamefont {Keppens}}, \bibinfo {author} {\bibfnamefont {C.~T.}\
			\bibnamefont {Nelson}}, \bibinfo {author} {\bibfnamefont {A.~F.}\
			\bibnamefont {May}}, \bibinfo {author} {\bibfnamefont {M.}~\bibnamefont
			{Brahlek}}, \bibinfo {author} {\bibfnamefont {C.}~\bibnamefont {Mazzoli}},
		\bibinfo {author} {\bibfnamefont {J.}~\bibnamefont {Pelliciari}}, \bibinfo
		{author} {\bibfnamefont {V.}~\bibnamefont {Bisogni}}, \bibinfo {author}
		{\bibfnamefont {V.~R.}\ \bibnamefont {Cooper}},\ and\ \bibinfo {author}
		{\bibfnamefont {T.Z.}\ \bibnamefont {Ward}},\ }\bibfield  {title} {\bibinfo
		{title} {Variance induced decoupling of spin, lattice, and charge ordering in
			perovskite nickelates},\ }\href
	{https://doi.org/10.1103/PhysRevResearch.5.013008} {\bibfield  {journal}
		{\bibinfo  {journal} {Phys. Rev. Res.}\ }\textbf {\bibinfo {volume} {5}},\
		\bibinfo {pages} {013008} (\bibinfo {year} {2023})}\BibitemShut {NoStop}%
\end{thebibliography}
\end{document}